\documentclass[12pt]{article}

\pdfoutput=1
\usepackage{amsmath,amssymb}
\usepackage{tikz}
\usepackage{graphicx}
\usepackage{appendix}

\parskip=\medskipamount

\numberwithin{equation}{section}

\title{Analytic Methods to Find Beating Transitions of Asymmetric Gaussian Beams in GNLS equations}
\author{David Ianetz$^{1,2,3}$ and Jeremy Schiff$^{1,4}$\\
   $^1$Department of Mathematics, Bar-Ilan University, Ramat-Gan 5290002, Israel\\
   $^2$Holon Institute of Technology (HIT), Holon 5810201, Israel \\ 
  $^3$E-mail: David.Ianetz@biu.ac.il\\
  $^4$E-mail: schiff@math.biu.ac.il}

\date{\today}

\begin{document}
\maketitle
\begin{abstract}
  In a simple model of propagation of asymmetric Gaussian beams in nonlinear waveguides,
  described by a reduction to ordinary differential eqautions of
  generalized nonlinear Schr\"odinger equations (GNLSEs)   
  with cubic-quintic (CQ) and saturable (SAT) nonlinearities and a graded-index profile,
  the beam widths exhibit two different types of beating behavior, with transitions between them.
  We present an analytic model to explain these phenomena, which originate in a $1:1$ resonance
  in a 2 degree-of-freedom Hamiltonian system. We show how small oscillations near a fixed
  point close to $1:1$ resonance in such a system can be approximated using an integrable
  Hamiltonian and, ultimately, by a single first order differential equation. In particular, the
  beating transitions can be located from coincidences of roots of a pair of
  quadratic equations, with coefficients determined (in a highly complex manner) by the
  internal parameters and initial conditions of the original system. The results of the analytic
  model agree with numerics of the original system over large  parameter ranges, and allow
  new predictions that can be verified directly. In the CQ case we identify a band of beam energies
  for which there is only a single beating transition (as opposed to $0$ or $2$) as the eccentricity is increased.  
  In the SAT case we explain the sudden (dis)appearance of beating transitions for certain
  values of the other parameters as the grade-index is changed.  
\end{abstract}


\newpage 

\section{Introduction}
In the sequence of papers \cite{Ianetz2010PRA_1,Ianetz2010PRA_2,Ianetz2013PRA}
a variational approach was taken to investigate the propagation of
asymmetric (elliptic) Gaussian beams in nonlinear waveguides, with cubic-quintic and saturable nonlinearities and
a parabolic graded-index (GRIN) profile, as described by suitable 
generalized nonlinear Schr\"odinger equations (GNLSEs). 
The beam widths in the two transverse directions to the direction
of propagation were found to obey a set of ordinary differential equations which can be identified as 
the equations of motion of a point particle in certain rather complicated, but
tractable, 2d potentials. 
Numerical analysis of these equations revealed ``beating'' phenomena: in addition to fast oscillations, the beam
widths exhibit a (relatively) slow periodic variation. Furthermore, two types of beating were identified: In type I beating
the amplitude of oscillation of the beam width in one direction remains greater than the amplitude of oscillation in the
other direction, whereas in type II, there is an interchange between the widths in the two transverse directions. The type
of beating depends on the parameters of the system and initial eccentricity of the beam. Remarkably, as the initial
eccentricty or other parameters are changed, there can be a transition between types, and this transition
is characterized by a singularity in the ratio of the periods of the beating and of the fast oscillatory motion. 

The intention of the current paper is to provide a theoretical analysis of the beating phenomena and, in particular,
to present an approximate analytic method to find the transitions between types. The relevant tool is the analysis
of small oscillations in
$2$ degree-of-freedom Hamiltonian systems near a fixed point which is close to $1:1$ resonance. The fact that resonance
is the source of ``beating'' or ``energy transfer'' phenomena in mechanical systems is well known. A classic example
can be found in the paper of Breitenburger and Mueller \cite{BM1} on the elastic pendulum, which the authors describe as
a ``paradigm of a conservative, autoparametric system with an internal resonance''. The paper \cite{BM1} has other features in
common with our work (such as the use of action-angle variables and the fact that the analytic approximation used is 
a single elliptic function equation) but it is in the much simpler context of $1:2$ resonance. For other examples
of autoparametric resonance see, for example, \cite{vpapr,Hallerbook}. The most widely used tool for analysis of
systems near resonance is
the mutliple time scale method, see for example \cite{kc,ManMan} for  thorough presentations and many examples. For a typical
modern application see \cite{VS1,VS2}. However, averaging techniques present an alternative \cite{SVM}, and in
the context of Hamiltonian systems, working in action-angle coordinates has substantial advantages \cite{gensap}. A typical
study of a system near resonance will involve looking at the bifurcations of special solutions. In this context much attention
has been paid to the definition and identification of {\em nonlinear normal modes} --- see \cite{mikhbar} for a review, and
\cite{RPV} for an example in the context of $1:1$ resonance. 

The $2$ degree-of-freedom Hamiltonian systems we study have a discrete symmetry, and are approximated by a family of systems 
with  $1:1$ resonance studied  nearly $40$ years ago by Verhulst \cite{Verhulst1}. 
Verhulst showed the existence of an approximate second integral and used this to study bifurcations of special solutions
and their stability. Our work differs from that of Verhulst and other works on $1:1$ resonance in several regards. The 
bifurcation question we pose depends not only on the internal parameters of the system, but also on the initial conditions.
The question is not only one of identifying different types of solutions of the system, but also seeing how the type of solution changes
as both the initial condition and internal system parameters are varied. We have not seen a similar study in the
highly complex context of $1:1$ resonance.  Our methodology uses 
action-angle variables and canonical transformations (though in an appendix we show how to apply standard two time scale
techniques). Unlike in most existing studies, it is necessary
to compute the relevant canonical transformation to {\em second} order.
However, this does not affect the result that once the correct canonical transformation has been applied,
the resulting approximating Hamiltonian depends only on a single combination of the angle variables
and is integrable. The equations of motion for the integrable Hamiltonian can be reduced to a single 
first order differential equation, and the rich bifurcation structure of the systems we study can
reduces to understanding the bifurcations of roots of a pair of quadratic polynomials,
with coefficients that depend (in a complex, nonexplicit manner) on the internal parameters
of the systems and the initial conditions. 
Comparison with numerical results shows our method gives high-quality results in a significant region of
parameter space, and allows a variety of interesting new predictions. 

The structure of this paper is as follows.
In the next section we review the relevant models from nonlinear optics and the
collective variable approximation to obtain equations for the propagation of beam widths,
and present the main findings of papers \cite{Ianetz2010PRA_1,Ianetz2010PRA_2,Ianetz2013PRA}  and some
further numerical results.  In section 3, we develop our 
method of integrable approximation for small oscillations in a $2$ degree-of-freedom Hamiltonian system 
near a fixed point close to $1:1$ resonance.
In section 4 we describe the application of this method to the specific systems relevant to beam propagation,
confirming existing numerical results and presenting new predictions. 
In section 5 we summarize and conclude. Appendix A completes some technical details omitted from the main text, 
and Appendix B describes an alternate
method of approximation of the full equations using a two time expansion. This is a more {\em ad hoc} approach than the one
explained in section 3, but we include it as it is more commonly used in the literature, 
and for certain values of parameters it gives better results. 

Before closing this introduction we mention a number of points concerning the relevance of the work
in this paper to optical solitons. We will describe in next section the manner in which we use ordinary differential
equations (ODEs) to study the behavior of solutions of GNLSEs. The use of ODEs to study GNLSEs is widespread, see for example 
   \cite{   Skarka2006PRL,Skarka2008JOA,Skarka2012PS,hemalomed,Skarka2014PRA,Aleksic2015PRA}
   In particular, the last two papers use ODE methods in the study of rotating solitons.
Our work extends the catalog of interesting bifurcations that can be observed in the context of GNLSEs; 
for another example; see the papers \cite{gmc1,gmc2} for a case of a saddle-loop bifurcation. 
Finally, we mention that we neglect dispersive terms in the GNLSEs we study. 
  This is justifiable in the context of  new optical materials
  \cite{Kim2002OL,Moon2008JNCS,JU2011OE,gold,noptrev}
characterized by Kerr coefficients of the order $10^{-11}$--$10^{-12}$ ${\rm cm}^2/{\rm W}$, making
the critical intensity for self-focusing  small enough that it can be reached
using microsecond pulses and possibly even continuous wave (CW) laser beams. 

\section{Models, the collective variable approach and numerical results}

We consider beam propagation in a nonlinear, graded-index fiber, as described by one of the following GNLSEs:
\begin{eqnarray}
  2i\psi_z + \psi_{xx} + \psi_{yy} + \left(  |\psi|^2 - Q|\psi|^4 - g(x^2+y^2) \right) \psi &=& 0\ ,
     \label{glnse1}\\  
 2i\psi_z + \psi_{xx} + \psi_{yy} + \left(  \frac{|\psi|^2}{1+\alpha^2|\psi|^2}  - g(x^2+y^2) \right) \psi &=& 0\ .
        \label{glnse2}  
\end{eqnarray}
Here, modulo suitable normalizations \cite{Kivshar2003book,Ianetz2013PRA},
$\psi$ is the strength of the electric field, $z$ is the
longitdinal coordinate, $x,y$ are transverse coordinates, and $Q,\alpha,g$ are parameters. 
The first equation is the case of cubic-quintic nonlinearity (CQ), the second is the case of saturable nonlinearity (SAT).
In the low intensity limit these models are similar, but for higher intensity they display different physical properties. 
In both cases, the higher order nonlinearity prevents beam collapse associated
with the standard Kerr nonlinearity  \cite{Chen2004PRE,Kivshar2003book}.
The term $-g(x^2+y^2)\psi$ reflects the graded-index nature of the fibre,
that the refractive index $n$ falls
with distance $r$ from the center of the fibre according to the law $n^2 = n_0^2 - Gr^2$; the physical significance of this 
is explained in \cite{Sodha1977book,Ghatak1978book,Ianetz2013PRA}.   

The collective variable approximation (CVA), introduced for the study of self-focusing beams in
\cite{Anderson1979PF,Anderson1979PF2,Anderson1983PRA}, is a variational technique
to approximate solutions of nonlinear Schr\"odinger-type equations which has been
used and validated in many different situations \cite{Malomed2002PO}. 
The method replaces partial differential equations such as (\ref{glnse1}) and (\ref{glnse2}) by
a system of ordinary differential equations
for the coefficients of an ansatz for the full solution.
The GNLSEs (\ref{glnse1}) and (\ref{glnse2}) are variational equations for action principles based on
the Lagrangian densities
\begin{eqnarray}
  {\cal L}_{\rm CQ} &=& i\left( \psi\psi^*_z-\psi^*\psi_z \right) + \left|\psi_{x}\right|^2 + \left|\psi_{y}\right|^2 
       - \frac12 |\psi|^4  + \frac{Q}{3}|\psi|^6   
           + g(x^2+y^2) |\psi|^2 \ ,  \label{lag1}\\
  {\cal L}_{\rm SAT} &=&  i\left( \psi\psi^*_z-\psi^*\psi_z \right) + \left|\psi_{x}\right|^2 + \left|\psi_{y}\right|^2 
  + \frac{\ln\left(1 + \alpha^2|\psi|^2\right)-\alpha^2|\psi|^2}{\alpha^4}  
  + g(x^2+y^2) |\psi|^2\ . \nonumber \\
    && \label{lag2}
\end{eqnarray}
We assume $\psi$ takes the form of the {\em trial function} 
\begin{equation}
  \label{trial_function}
  \psi_{T}(x,y,z) = A(z)
  \exp \left(i\phi(z)
  -\frac {x^2} {2a_x^2(z)} + ib_x(z)x^2
  -\frac {y^2} {2a_y^2(z)} + ib_y(z)y^2
  \right)\ ,
\end{equation}
where $A,\phi,a_x,a_y,b_x,b_y$ are currently undetermined, real functions of only the longitudinal coordinate $z$.
This trial function describes an elliptic Gaussian
beam with $a_x,a_y$ representing the widths of the beam in the $x,y$ directions. $b_x,b_y$ describe curvatures
of the beam wavefront, $A$ is the normalized amplitude of the electric field, and $\phi$ is a longitudinal phase factor.
Our choice of a Gaussian shape for the trial function is appropriate because the Gaussian is an exact
solution of the linear Schr\"odinger equation for GRIN waveguides \cite{Sodha1977book,Ghatak1978book}.
Substituting the trial function in the Lagrangian densities
(\ref{lag1}),(\ref{lag2}) and computing the integrals over the variables $x,y$ we obtain reduced densities
for the functions $A,\phi,a_x,a_y,b_x,b_y$. The corresponding Euler-Lagrange equations
in the CQ case are 
\begin{eqnarray*}
     \dot{A}&=& -(b_x +b_y)A \ , \\
     \dot{a}_{x,y} &=& 2a_{x,y}b_{x,y}\ ,  \\
     \dot{b}_{x,y} &=& \frac{1}{2a^4_{x,y}} - 2b^2_{x,y} - \frac{g}{2}
                - \frac{A^2}{a^2_{x,y}}\left( \frac{1}{8}- \frac{QA^2}{9} \right)\ ,  \\
     \dot{\phi} &=& - \frac{1}{2a^2_x} - \frac{1}{2a^2_y} + \left( \frac{3}{8} - \frac{5QA^2}{18} \right) A^2  
                 \ . 
\end{eqnarray*}
Here a dot denotes differentiation with respect to $z$. 
In the SAT case the equations for $A,a_x,a_y$ remain the same, but those for $b_x,b_y,\phi$  are replaced by 
\begin{eqnarray*}
     \dot{b}_{x,y} &=& \frac {1} {2a_{x,y}^4} -2b_{x,y}^2 - \frac {g} {2} +
                       \frac {\ln(1+ \alpha^2 A^2)+\textrm{Li}_{2}(-\alpha^2 A^2)}{2\alpha^4 A^2a_{x,y}^2}\ , \\
                       \dot{\phi} &=&  -\frac{1}{2a_x^2} - \frac{1}{2a_y^2}
                       + \frac{\alpha^2A^2-2\ln(1+\alpha^2A^2)- \textrm{Li}_2(-\alpha^2 A^2)}{2\alpha^4A^2}\ .
\end{eqnarray*}
Here $\textrm{Li}_2(x)=\sum_{k=1}^{\infty} \frac{x^k}{k^2}$  is the Spence or dilogarithm function  \cite{Abramowitz1970book}. 
For both CQ and SAT cases we observe that $A^2 a_xa_y$ is conserved \cite{Ianetz2010PRA_1,Ianetz2010PRA_2,Ianetz2013PRA},
and we write $A^2 a_x a_y = 4 E$ ($4E$ is the beam energy), and use this to
eliminate $A(z)$ . Furthermore, $\phi(z)$ evidently plays no role in determining the other functions and
can be computed by a simple quadrature once the other functions have been found. 
Furthermore, it is clear that
we can write $b_x$ ($b_y$) in terms of $a_x$  ($a_y$) and its $z$-derivative. Thus we can reduce the system of $6$
equations to a pair of second order equations for $a_x,a_y$. After some more calculation it emerges that the
equations are simply the equations of motion 
\begin{equation}
    \ddot{a}_x = -\frac{\partial V}{\partial a_x}  \ , \qquad
    \ddot{a}_y = -\frac{\partial V}{\partial a_y}
\label{em}\end{equation}    
for a particle in a potential $V(a_x,a_y)$, where for CQ
\begin{equation}
V = V_{CQ} \equiv  \frac12 \left( \frac1{a_x^2} + \frac1{a_y^2} \right) 
- \frac{E}{a_x a_y}
+ \frac{16 QE^2}{9a_x^2a_y^2} 
+ \frac{g}{2}   \left( a_x^{2}+a_y^{2} \right)\ ,
\label{VCQ}\end{equation}
and for SAT 
\begin{equation}
V = 
V_{SAT} \equiv  \frac12 \left( \frac1{a_x^2} + \frac1{a_y^2} \right) 
-\frac{a_xa_y}{4E\alpha^4} {\rm Li}_2 \left( -\frac{4E\alpha^2}{a_xa_y} \right) 
+ \frac{g}{2}   \left( a_x^{2}+a_y^{2} \right)\ . 
\label{VSAT}\end{equation}

Thus integration of equations (\ref{em}) for potentials (\ref{VCQ}) and (\ref{VSAT}) provides a
first approximation to solutions of the GNLSEs (\ref{glnse1}) and (\ref{glnse2}). Full numerical solutions
of  GNLSEs have been given in both 
the SAT \cite{Yang2002PRE} and the CQ \cite{Michinel2002PRE} cases with $g=0$. 
In  \cite{Michinel2002PRE} it was shown that the breathing frequencies found numerically are
similar to those obtained by the CVA technique.
In \cite{Yang2002PRE} it was shown that 
the shape of the beam obtained numerically for a saturable medium remains similar to  Gaussian, even
for an asymmetric initial condition. 
However, use of direct numeric methods
to give an overall picture of the behavior of a GNLSE, as a function of all the various parameters, 
remains a computationally overwhelming task, and
having an qualitatively correct analytic or semianalytic model is therefore
useful for developing physical insight
\cite{Kivshar2003book,Agrawal2007book}. 

Appropriate initial conditions for (\ref{em}) are
\begin{equation} 
a_x(0) = a_0 r \ , \qquad  a_y(0) = \frac{a_0}{r} \ , \qquad
\dot{a_x}(0) = \dot{a_y}(0) = 0\ .   \label{ic0}\end{equation}
The latter two conditions are equivalent to taking $b_x(0)=b_y(0)=0$. 
Note that both the CQ and the SAT system have a scaling symmetry
\begin{equation}
\begin{array}{lllll} 
a_x \rightarrow \lambda a_x \ , &
a_y \rightarrow  \lambda a_y\ , &
a_0 \rightarrow  \lambda a_0 \ , & 
r   \rightarrow r \ , &
z \rightarrow  \lambda^2 z \ ,\\ 
Q  \rightarrow   \lambda^2  Q\ ,  &
\alpha  \rightarrow   \lambda  \alpha\ ,  &
g  \rightarrow   \lambda^{-4}  g\ ,   &
E  \rightarrow   E  .  &
\end{array}
\end{equation}
Thus for CQ we do not need to study the dependence of solutions on the $5$ parameters $Q,g,E,a_0,r$, but only on
the $4$ scale invariant quantities $Qa_0^{-2}, ga_0^4, E, r$. On occasion we will work 
with the scale invariant quantity  $K_{\rm CQ} = 4QEa_0^{-2}$ instead of the quantity $Qa_0^{-2}$.
(For SAT, replace all instances of $Q$ in the previous two sentences with $\alpha^2$, and $K_{\rm SAT}=4\alpha^2Ea_0^{-2}$.)
Note that since there is symmetry in both the models between $a_x$ and $a_y$, there is a $r\rightarrow \frac1{r}$ inversion symmetry,
and thus we need only study $r\le 1$ or $r\ge 1$. 
   
In the papers \cite{Ianetz2010PRA_1,Ianetz2010PRA_2,Ianetz2013PRA} the ODE systems above were studied numerically.
For appropriate choices of the parameters ``beating'' phenomena were observed: in addition to (relatively) fast
``breathing'' oscillations, the beam
widths exhibit a (relatively) slow periodic variation. Two types of  beating were identified: 
In type I beating,
the amplitude of oscillation of the beam width in one direction remains greater than the amplitude of oscillation in the
other direction. In type II beating, there is an interchange between the widths in the two transverse directions. 
This is illustrated in Figure 1, which shows solutions of the CQ system
for $E=2.039$, $K_{\rm CQ}=0.71$, $ga_0^4=0.01$, and two choices of $r$: $r=1.14$ gives type I beating, whereas 
$r=1.16$ gives type II beating.

\begin{figure}[t]
\center{
  \includegraphics[width=0.98\textwidth]{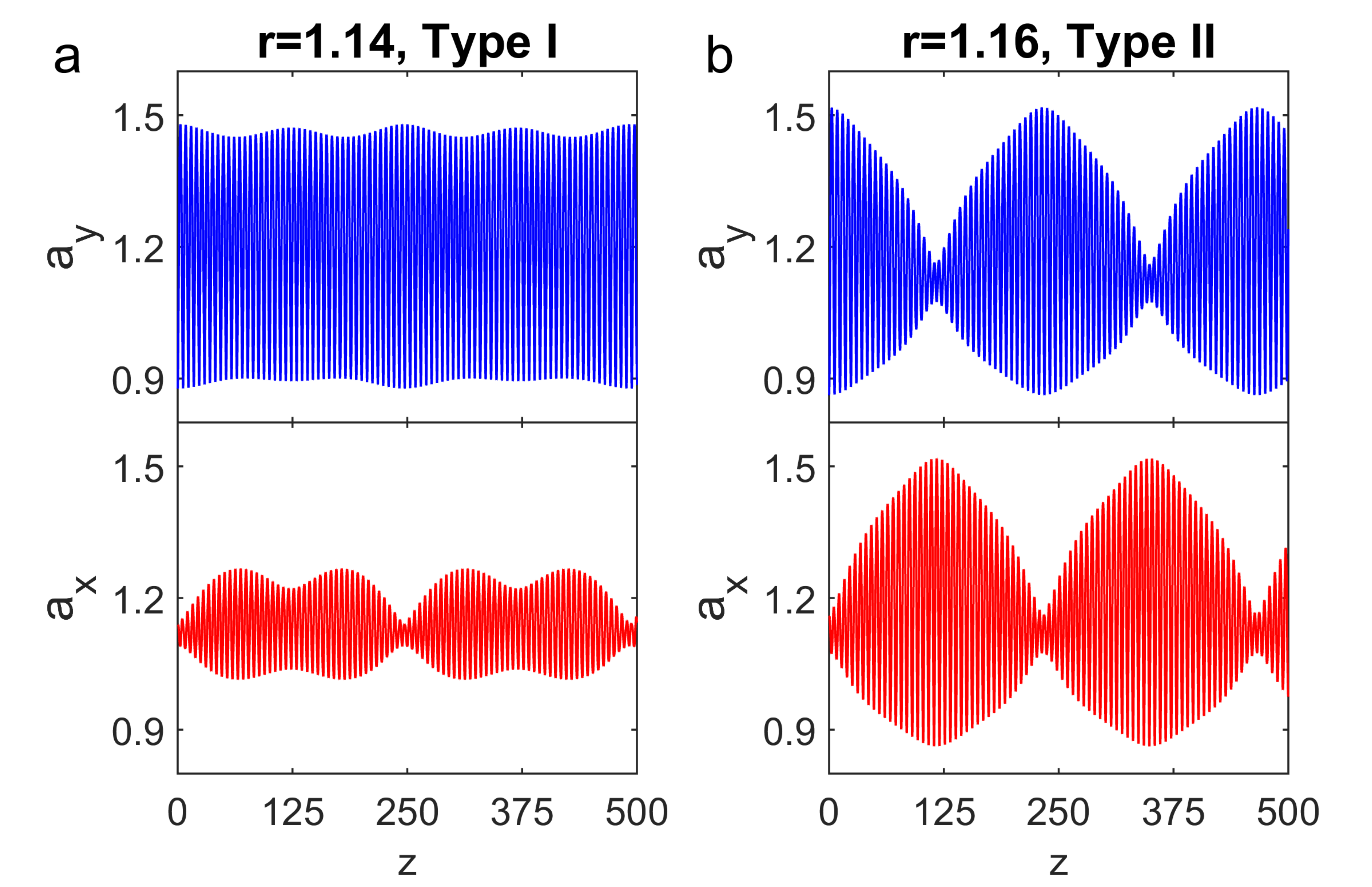}
}
\caption{\label{beating}
  Two types of beating in the CQ system, with $E=2.039$, $K_{\rm CQ}=0.71$, $ga_0^4=0.01$. (a) For $r=1.14$ the amplitude of
  oscillation of $a_y$ is always more than that of $a_x$, this is type I beating. (b) For $r=1.16$ there is an exchange of roles,
  this is type II beating. 
  }
\end{figure}

\begin{figure}[t]
\center
\includegraphics[width=0.7\textwidth]{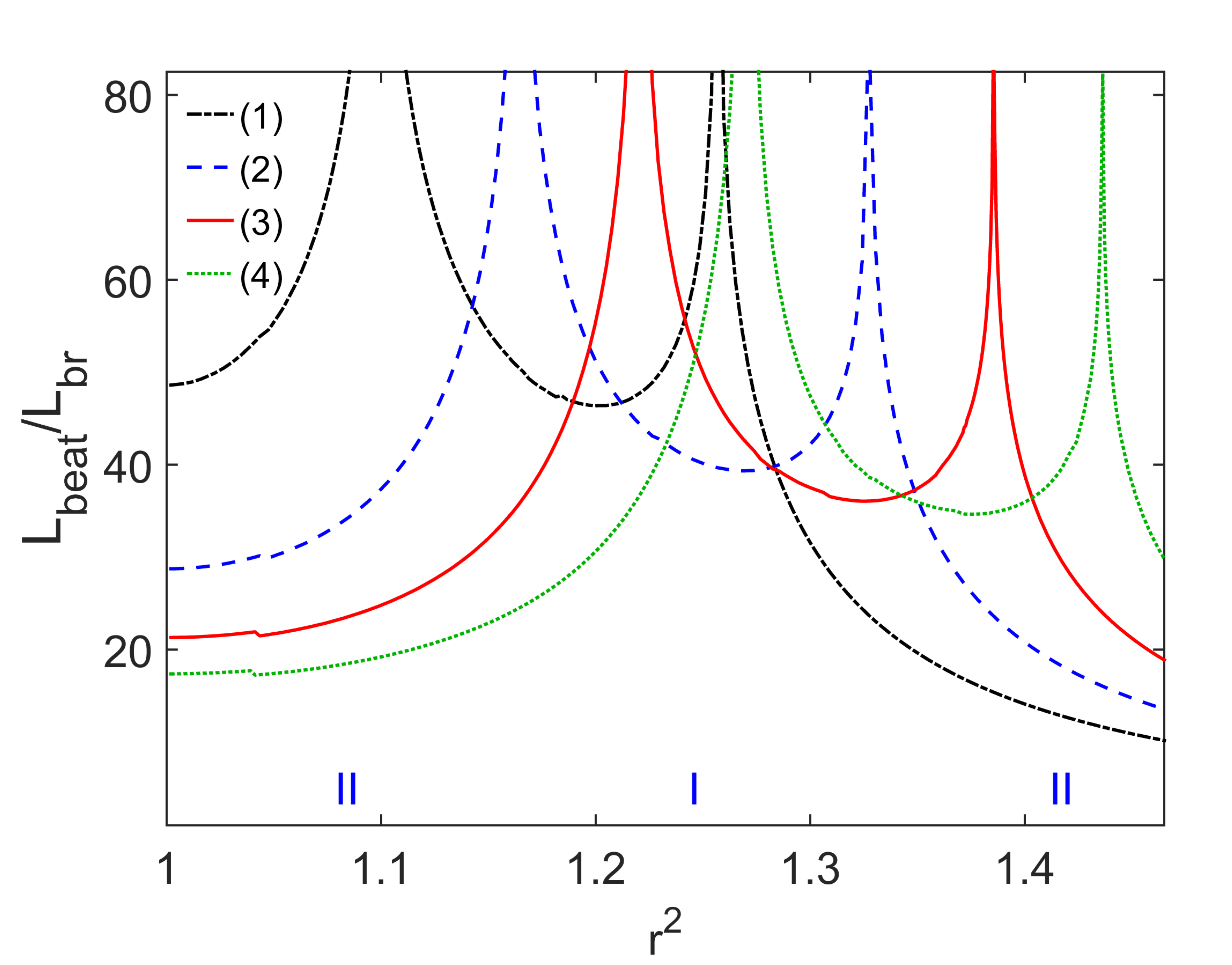}
\caption{\label{fig:epsart} Dependence of the ratio of periods of slow beating and fast oscillatory motion, 
  $L_{\rm beat}/L_{\rm br}$, on the parameter $r^2$ for the CQ model, and for various values of the parameter $ga_0^4$. 
  $E=2.039$ and $K_{\rm CQ}=0.71$ throughout. 
  (1) (dash-dot black)  $ga_0^4=0$, (2) (dashed blue)  $ga_0^4=0.01$, (3) (solid red)  $ga_0^4=0.02$,
  (4) (dotted green)  $ga_0^4=0.03$. Roman numerals indicate the type of beating in case (2), $ga_0^4=0.01$.}
\end{figure}

The type of beating depends on the parameters of the system and, as evident from Figure 1, on the  initial eccentricity of the beam.
Remarkably, as the initial eccentricty is increased, or as other parmeters are changed, there can be a transition between  
types. The approach to this transition is characterized by a divergence in the ratio of the periods of the slow beating and of
the fast oscillatory motion. In Figure 2 this ratio (determined from numerical simulations) is plotted as a
function of $r^2$ for the CQ system,
with $E=2.039$, $K_{\rm CQ}=0.71$ and  $ga_0^4=0,0.01,0.02,0.03$. (The reason for the choice of the coordinate $r^2$ on the $x$-axis is
simply to make the plot clearer.)
For $r$ just above
$1$ the beating is type II, then there is a transition to type I, and then a second transition back to type II.
The dependence on the system parameters
of the two critical values of $r$, which we denote collectively by $r_c$, 
is explored further in Figure 3. 
In Figure 3a  the values of $r_c$ are plotted as a function of $ga_0^4$ for three different values of
$K_{\rm CQ}$ and a constant value of  $E$;
in Figure 3b $r_c$ is plotted as a function of $ga_0^4$ for three different values of
$E$ and a constant value of  $K_{\rm CQ}$.
In general we see that $r_c$ increases as a function of $ga_0^4$ (for fixed $E,K_{\rm CQ}$). 
From Figure 3b we see that since the (solid) red is above the (dashed) blue is above the (dot-dashed) black,
$r_c$ also increases as a function of $E$ (for fixed $ga_0^4,K_{\rm CQ}$). But in Figure 3a we see there is
difference between the upper  and lower branches of $r_c$. We deduce that 
the higher value of $r_c$ also increases with $K_{\rm CQ}$ (for fixed $E,ga_0^4$), but the
lower value decreases. 

\begin{figure}[t]
\center{
\includegraphics[width=0.98\textwidth]{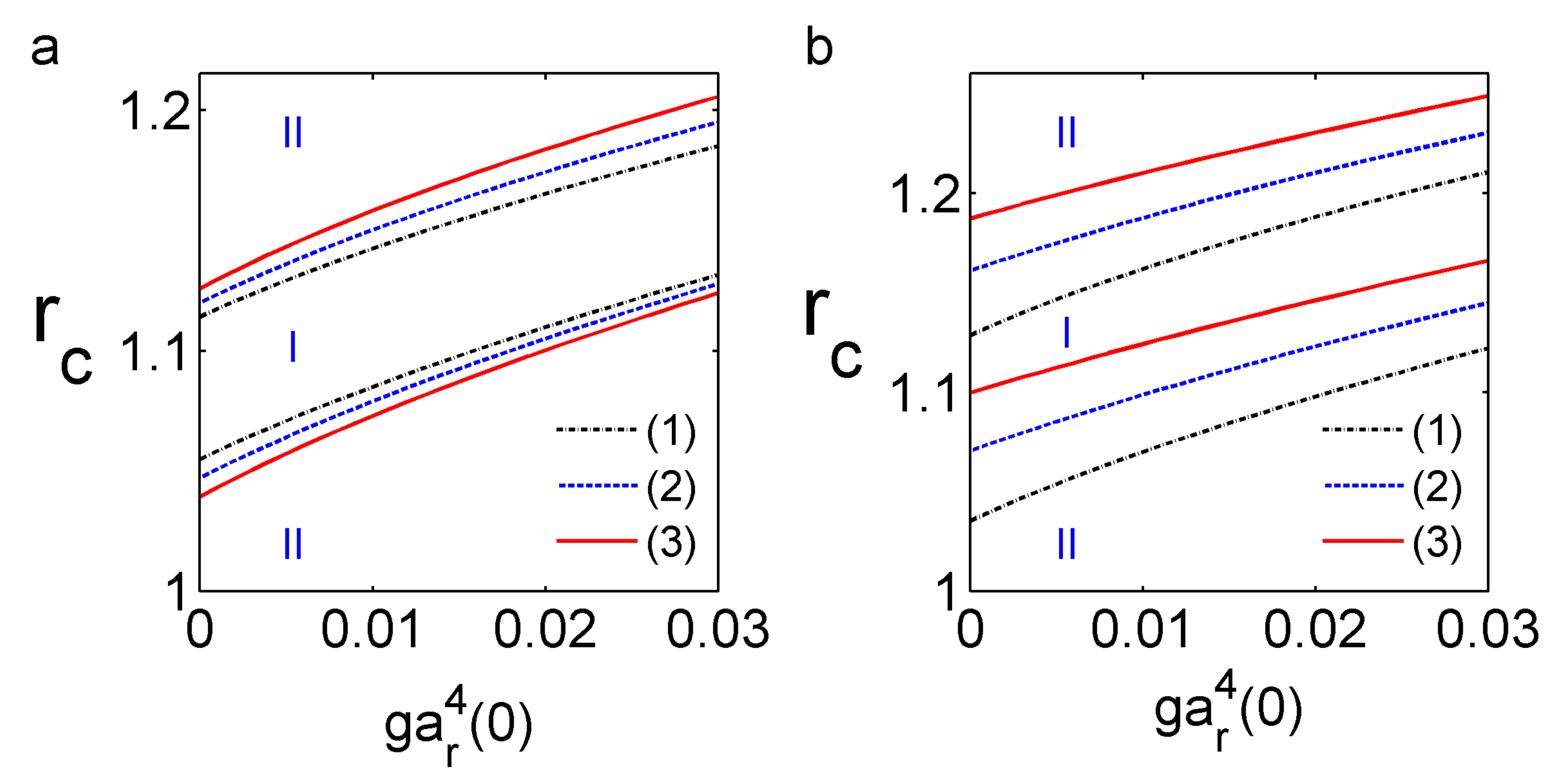}
}
\caption{\label{critEs}
  Dependence of $r_c$, the critical values of $r$, on system paramters $ga_0^4,K_{\rm CQ},E$.
  (a) $r_c$ as a function of $ga_0^4$ for 
  (1) $K_{\rm CQ}=0.602$ (dash-dot black), (2) $K_{\rm CQ}=0.71$ (dashed blue) and (3) $K_{\rm CQ}=0.739$ (solid red), 
  all with $E=2.039$. 
  (b) $r_c$ as a function of $ga_0^4$ for 
  (1) $E=2.039$ (dash-dot black), (2) $E=2.079$ (dashed blue) and (3) $E=2.12$ (solid red), all with
  $K_{\rm CQ}=0.753$. 
}
\end{figure}


We shall see later that for other values of $ga_0^4,K_{\rm CQ},E$ there can be just a single transition or
no transitions at all as $r$ is increased from $1$.  Transitions between beating types are also
observed in the SAT system, again with a complex dependence on the parameters $ga_0^4$, $K_{\rm SAT}$ and $E$. 
The aim of this paper is to provide an integrable approximation for 
equations (\ref{em}) with  potentials (\ref{VCQ}) and (\ref{VSAT}) which provides
a theoretical model to predict where the transitions between types take place.   

\section{Small oscillations near $1:1$ resonance}


In this section we describe a general process of approximation near a $1:1$ resonance 
for a $2$ degree-of-freedom Hamiltonian system with Hamiltonian  
\begin{equation}
H =  \frac12 \left(p_x^2 + p_y^2 \right)  + V(a_x,a_y)\ . 
\label{H}\end{equation}
Here $a_x,a_y$ are the coordinates, $p_x,p_y$ are the conjugate momenta, and the potential $V$ (which typically will depend on
a number of parameters) is symmetric, $V(a_x,a_y)=V(a_y,a_x)$.
We assume that for typical values of the parameters 
the potential has an isolated symmetric minimum (at $a_x=a_y=a_{\rm min}$, say) at which the system is close to $1:1$ resonance.
Note that because of the symmetry, $\displaystyle{  \frac{ \partial^2V}{\partial a_x^2} (a_{\rm min}, a_{\rm min})
  = \frac{ \partial^2V}{\partial a_y^2} (a_{\rm min},a_{\rm min})}$. 
Thus the Hessian matrix of the potential at $(a_{\rm min},a_{\rm min})$ has
eigenvectors $\left( \begin{array}{c} 1 \\ \pm 1  \end{array}  \right)$
with eigenvalues $\displaystyle{\frac{ \partial^2V}{\partial a_x^2} (a_{\rm min},a_{\rm min})
  \pm  \frac{ \partial^2V}{\partial a_xa_y} (a_{\rm min},a_{\rm min})}$.
The condition for being close to $1:1$ resonance (i.e. equal eigenvalues) is therefore simply
$  \frac{ \partial^2V}{\partial a_xa_y} (a_{\rm min},a_{\rm min}) \approx 0 $. 
For these systems we study orbits with initial conditions as given in (\ref{ic0}).

The process of approximating such a system with an integrable  system has $3$ steps. 

The {\em first step} is to expand in normal coordinates near the fixed point, retaining only
terms up to order $4$ in the potential. Thus we write
$$  a_x = a_{\rm min} + \frac{\zeta_2+\zeta_1}{\sqrt{2}} \ , \qquad   a_y = a_{\rm min} + \frac{\zeta_2-\zeta_1}{\sqrt{2}}  $$
and expand to fourth order to obtain
\begin{equation}
H_1 = \frac12 \left( p_1^2 + p_2^2 + \omega_1^2 \zeta_1^2 + \omega_2^2 \zeta_2^2 \right)  + a_1 \zeta_1^2 \zeta_2 + a_2 \zeta_2^3
+ a_3 \zeta_1^4 + a_4 \zeta_1^2 \zeta_2^2 + a_5 \zeta_2^4
\label{H1}\end{equation}
where $p_1,p_2$ are the conjugate momenta to the coordinates $\zeta_1,\zeta_2$, and $\omega_1,\omega_2,a_1,a_2,a_3,a_4,a_5$
are constants that depend on the parameters of the original potential $V$. The Hamiltonian $H_1$ has $\zeta_1 \rightarrow -\zeta_1$
symmetry as a consequence of the symmetry of $H$, and is the general Hamiltonian with this symmetry and a quartic potential. 
Aspects of the behavior of this Hamiltonian at, or close to, $1:1$ resonance have been studied previously, for example,  in 
\cite{Verhulst1,montaldi1,montaldi2,pm3,pm1,pm2}. The initial conditions
for this system, corresponding to (\ref{ic0}) are 
\begin{equation} 
  \zeta_1(0) = \frac{a_0}{\sqrt{2}} \left(r-\frac1{r} \right)
  \ , \qquad  \zeta_2(0) = \frac1{\sqrt{2}} \left(a_0\left(r+\frac1{r} \right) - 2 a_{\rm min} \right)
  \ , \qquad
   {p_1}(0) = {p_2}(0) = 0 \ .  \label{ic1}\end{equation} 
Symmetric solutions correspond to the initial condition $\zeta_1(0)=0$. In regarding $H_1$ as an approximation for $H$
we are neglecting terms of fifth order and above.  

The {\em second step} is to make the canonical
transformation to action-angle coordinates associated with the quadratic part of the Hamiltonian $H_1$,
i.e. to substitute
$$
\begin{array}{ll}
  \zeta_1 = \sqrt{\frac{2J_1}{\omega_1}} \cos\theta_1\ ,  &   p_1 = - \sqrt{2J_1\omega_1} \sin\theta_1 \ ,   \\
  \zeta_2 = \sqrt{\frac{2J_2}{\omega_2}} \cos\theta_2\ ,  &   p_2 = - \sqrt{2J_2\omega_2} \sin\theta_2    \ .   
\end{array}
$$
This gives 
\begin{eqnarray}
  H_2 &=&  \omega_1 J_1 + \omega_2 J_2  
  +   \left( \frac{2J_1a_1}{\omega_1} +\frac{3 J_2 a_2}{\omega_2} \right) \sqrt{\frac{J_2}{2\omega_2}} \cos\theta_2  \nonumber\\
 && 
  + \left(\frac{J_2}{2\omega_2}\right)^{3/2}  2a_2\cos 3\theta_2
  +  a_1 \frac{J_1}{\omega_1} \sqrt{\frac{J_2}{2\omega_2}}
     \left( \cos(2\theta_1-\theta_2)  + \cos(2\theta_1+\theta_2) \right)    \nonumber\\ 
&& +   3a_3 \frac{J_1^2}{2\omega_1^2}
     + a_4 \frac{J_1J_2}{\omega_1\omega_2}
     + 3a_5 \frac{J_2^2}{2\omega_2^2}
   + a_4 \frac{J_1J_2}{2\omega_1\omega_2}\left( \cos(2\theta_1-2\theta_2)  + \cos(2\theta_1+2\theta_2) \right)
     \nonumber  \\
&&  \left( 2a_3 \frac{J_1^2}{\omega_1^2} +  a_4 \frac{J_1J_2}{\omega_1\omega_2}  \right)\cos(2\theta_1)
        + a_3 \frac{J_1^2}{2\omega_1^2} \cos(4\theta_1)
        +\left(a_4\frac{J_1J_2}{\omega_1\omega_2}+ 2a_5\frac{J_2^2}{\omega_2^2} \right)\cos(2\theta_2)   \nonumber\\
&&         +a_5\frac{J_2^2}{2\omega_2^2}\cos(4\theta_2)   \ .  \label{H2}   
\end{eqnarray}
Here $\theta_1,\theta_2$ are the angle variables, and $J_1,J_2$ the conjugate actions. 
The initial conditions for the action variables are
\begin{equation} 
J_1(0) = \frac{a_0^2\omega_1}{4} \left(r-\frac1{r} \right)^2 \ , \qquad
J_2(0) =   \frac{\omega_2}{4} \left(a_0\left(r+\frac1{r} \right) - 2 a_{\rm min} \right)^2\ .
\label{Jic}
\end{equation}
The initial conditions for the angle variables depend on the sign of $\zeta_1(0)$ and $\zeta_2(0)$. 
If $\zeta_1(0)>0$ ($\zeta_2(0)>0$) then, from (\ref{ic1}) we should take $\theta_1(0)=0$ ($\theta_2(0)=0$)
and otherwise $\theta_1(0)=\pi$ ($\theta_2(0)=\pi$). Due to the $\zeta_1\rightarrow -\zeta_1$ symmetry of $H_1$ 
the Hamiltonian $H_2$ has period $\pi$ (and not $2\pi$) as a function of $\theta_1$ and thus the choice of
the $\theta_1$ initial condition is irrelevant. The choice of the $\theta_2$ initial condition, however, is
important. We are introducing a non-physical discontinuity in the approximation procedure when
the sign of $\zeta_2(0)$ changes, i.e. when  $r+\frac1{r}=\frac{2a_{\rm min}}{a_0}$. We will see the effects of this
later, in our results for the SAT potential. 

The {\em third step} involves a canonical change of coordinates $(\theta_1,\theta_2,J_1,J_2)\rightarrow (\phi_1,\phi_2,K_1,K_2)$
defined by a generating function of the second type $G_2(\theta_1,\theta_2,K_1,K_2)$ \cite{Goldstein}, chosen to eliminate the
nonresonant terms from the Hamiltonian (i.e. all the trigonometric terms of order $||J||^{3/2}$ or $||J||^2$ except the one involving
$\cos(2\theta_1-2\theta_2)$.) 
The full change of coordinates is given by 
\begin{equation}
\begin{array}{ll}  
  \phi_1 = {\displaystyle{\frac{\partial G_2}{\partial K_1}}} \ , &
    J_1 = {\displaystyle{\frac{\partial G_2}{\partial \theta_1}}} \ , \\[\bigskipamount]
      \phi_2 = {\displaystyle{\frac{\partial G_2}{\partial K_2}}} \ , &
        J_2 = {\displaystyle{\frac{\partial G_2}{\partial \theta_2}}} \ . 
\end{array}
\label{ct}\end{equation}
The generating function $G_2$ should be taken in the form
\begin{eqnarray*}
  G_2 &=&  K_1\theta_1 + K_2\theta_2 +  A_1 \sin\theta_2 + A_2\sin 3\theta_2 +  A_3 \sin(2\theta_1-\theta_2)
        + A_4 \sin(2\theta_1+\theta_2)  \\
        &&  +  A_5 \sin2\theta_1 + A_6 \sin 4\theta_1  + A_7 \sin2\theta_2 + A_8 \sin 4\theta_2
        + A_9 \sin6\theta_2  + A_{10} \sin (2\theta_1 +2\theta_2)\\
        &&  
        + A_{11} \sin (2\theta_1 +4\theta_2)
        + A_{12} \sin (4\theta_1 +2\theta_2)
        + A_{13} \sin (2\theta_1 -4\theta_2) 
        + A_{14} \sin (4\theta_1 -2\theta_2)  
\end{eqnarray*}
where the coefficients $A_1,\ldots,A_{14}$ are functions of $K_1,K_2$, which  are chosen to eliminate the nonresonant
trigonometric terms in the Hamiltonian to required order.  $A_1,A_2,A_3,A_4$ are of order $||K||^{3/2}$ and $A_5,\ldots,A_{14}$
are of order $||K||^2$.  The calculations are long, but straightforward with the help of a symbolic manipulator, and the final
Hamiltonian is found to be simply 
\begin{equation}
H_3 =  \omega_1 K_1 + \omega_2 K_2 
+  b_1 K_1^2 + b_2 K_1K_2 + b_3 K_2^2 
+ (b_4 K_1^2 + b_5 K_1K_2) \cos\left( 2(\phi_1-\phi_2)   \right) 
\label{H3}\end{equation}
where 
\begin{eqnarray}  
  b_1 &=& \frac{3a_{{3}}}{2\omega_{1}^{2}} 
  -   \frac{a_{{1}}^{2}\left( 8\omega_{{1}}^{2}-3\omega_{{2}}^{2} \right)}
          {4\omega_1^{2}\omega_2^{2} \left( 2\omega_1-\omega_2 \right)  \left( 2\omega_1 + \omega_2 \right)}
\nonumber\\ 
b_2 &=&  \frac{a_{{4}}}{\omega_{{1}}\omega_{{2}}} - \frac{3 a_1 a_2}{\omega_1\omega_2^3}
- \frac {2a_{{1}}^2}
        {\omega_{{1}}\omega_{{2}}  \left( 2\omega_{{1}}-\omega_{{2}} \right)  \left( 2\omega_{{1}}+\omega_{{2}} \right) } 
\nonumber\\
  b_3 &=&  \frac{3a_{{5}}}{2\omega_{{2}}^{2}} -  \frac{15 a_{{2}}^{2}}{4\omega_{{2}}^{4}}
\label{bs}\\
b_4 &=&  \frac { \left( \omega_{{2}}-\omega_{{1}} \right) a_{{1}}^{2} }
               {2\omega_{{1}}^{2}\omega_{{2}}^{2} \left( 2\omega_{{1}}-\omega_{{2}} \right) }
\nonumber\\          
b_5 &=&  \frac {a_{{4}}}{2\omega_{{1}}\omega_{{2}}}
- \frac{a_1a_2(4\omega_1^2-3\omega_1\omega_2-4\omega_2^2)}
{2\omega_{{1}}\omega_{{2}}^{3} \left( 2\omega_{{1}}-\omega_{{2}} \right)\left( 2\omega_{{1}}+\omega_{{2}} \right)}
   - \frac{a_1^2}{\omega_1^2\omega_2(2\omega_1-\omega_2)}
\ . \nonumber
\end{eqnarray}

{\em The Hamiltonian $H_3$ given in (\ref{H3}) is an integrable approximation of the original Hamiltonian $H$
  given in (\ref{H}).}  $H_3$ is a {\em normal form} for the ``natural'' Hamiltonian $H_1$ at or near $1:1$ resonance.
Note that in the case of exact resonance  $\omega_1=\omega_2$  the coefficient $b_4$ vanishes. Also close to resonance,  
the corresponding term in $H_3$ is of lower order than the other terms, and in  \cite{pm3,pm1,pm2} it is omitted. 
However, we choose to  retain it to avoid any assumption on the relative orders of magnitude of $|\omega_1 - \omega_2|$
and $||K||$.  The integrability of $H_3$ is evident, as it only depends on the modified angle variables
$\phi_1,\phi_2$ through the combination $\phi_1-\phi_2$. As a consequence the quantity $K_1+K_2$ is conserved,
in addition to the Hamiltonian itself. We denote the value of the Hamiltonian by ${\cal E}$ and the value of $K_1+K_2$
by $P$ (these should be computed from the system parameters and initial conditions). The full equations of motion are 
\begin{eqnarray}  
  \dot{\phi}_1 = \frac{\partial H_3}{\partial K_1}  
  &=& \omega_1 + 2 b_1 K_1 + b_2 K_2 +  (2 b_4 K_1 + b_5 K_2) \cos(2(\phi_1-\phi_2))   \ , \label{fe1}\\  
  \dot{\phi}_2 = \frac{\partial H_3}{\partial K_2} 
               &=& \omega_2 +  b_2 K_1 + 2 b_3 K_2 + b_5 K_1 \cos(2(\phi_1-\phi_2))  \ ,  \label{fe2}\\
  \dot{K}_1 = -\frac{\partial H_3}{\partial \phi_1} 
  &=& 2 K_1 (b_4 K_1 + b_5 K_2) \sin (2(\phi_1-\phi_2))   \ ,  \label{fe3}\\ 
  \dot{K}_2 = -\frac{\partial H_3}{\partial \phi_2}  
            &=& -2 K_1 (b_4 K_1 + b_5 K_2) \sin (2(\phi_1-\phi_2)) \ . \label{fe4}
\end{eqnarray}
Using the two conservation laws it is possible to eliminate $K_2$ and $\phi_1-\phi_2$ from the $K_1$ equation of motion
to get a single equation for $K_1$:
\begin{eqnarray}  
  \dot{K}_1^2  &=& - 4 
  ((b_1-b_2+b_3+b_4-b_5)K_1^2+((b_2-2b_3+b_5)P-\omega_2+\omega_1)K_1+b_3P^2+\omega_2P-{\cal E})  \nonumber \\  
  &&
  ((b_1-b_2+b_3-b_4+b_5)K_1^2+((b_2-2b_3-b_5)P-\omega_2+\omega_1)K_1+b_3P^2+\omega_2P-{\cal E})\ .\nonumber \\
    &&   \label{K1e}
\end{eqnarray}


Equation (\ref{K1e}) is a central result of this paper. 
To solve (\ref{K1e}) it is necessary to translate the initial conditions for
$J_1,J_2,\theta_1,\theta_2$
into initial conditions for $K_1,K_2$. This step requires details of the canonical tranformation. 
Due to their length, the full equations determining the initial values of $K_1,K_2$ are given in Appendix A (equations 
(\ref{Keq1})-(\ref{Keq2})). Note there are two cases depending on whether $\theta_2(0)$ is $0$ or $\pi$. Note also that 
{\em there  is no guarantee that these equations will have a solution with real, positive $K_1,K_2$}. 
In the case of the SAT system, for a certain range of parameter values we have experienced numerical problems with
the solution of (\ref{Keq1})-(\ref{Keq2}), specifically for initial values of 
$J_2$ close to zero, close to the jump from $\theta_2=0$ to $\theta_2=\pi$. However, typically there are values
of $K_1(0),K_2(0)$ close to the given values of $J_1(0),J_2(0)$. 

Once the initial values of $K_1,K_2$ have been computed, the values of the constants ${\cal E}$ and $P$ can
be found and equation (\ref{K1e}) can be solved. The right hand side of (\ref{K1e}) is the product of two
quadratic factors in $K_1$, with up to $4$ real roots, and typical solutions will be  oscillatory between two roots.
When there is a {\em double root} then there is the possibility of the period of the oscillation becoming {\em infinite},
marking a bifurcation in the solution. There are two ways that a double root can occur,  by the vanishing
of the discriminant of one of the quadratic factors, or by one of the roots of the first factor coinciding with one
of the roots of the second. The discriminants of the quadratic factors are
\begin{eqnarray}
\Delta_1 &=& 
    \left( (b_2+b_5)^2 - 4b_3 (b_1 + b_4) \right) P^2 
    + 2 \left( (b_2 - 2 b_3 + b_5) \omega_1   +  (-2b_1 + b_2 - 2 b_4 +  b_5) \omega_2  \right) P  \nonumber \\
    &&    + 4  (b_1 - b_2 + b_3 + b_4 - b_5){\cal E} + (\omega_1 - \omega_2) ^ 2 \ ,  \label{Del1}    \\
\Delta_2  &=& 
    \left( (b_2-b_5)^2 -4b_3 (b_1 - b_4) \right) P^2 
    + 2 \left( (b_2 - 2 b_3 - b_5) \omega_1   +  (-2b_1 + b_2 + 2 b_4 -  b_5) \omega_2  \right) P \nonumber \\
&&     + 4  (b_1 - b_2 + b_3 - b_4 + b_5){\cal E} + (\omega_1 - \omega_2) ^ 2 \ . \label{Del2}
\end{eqnarray}
A simple algebraic manipulation shows that 
the first factor and second factor have coincident roots if either $\Delta_3=0$ or $\Delta_4=0$,
where
\begin{eqnarray}
\Delta_3 &=&  b_3 P^2 + \omega_2  P - {\cal E}\ ,   \label{Del3} \\   
\Delta_4  &=&  \frac{ b_1 b_5^2 - b_2 b_4 b_5 + b_3 b_4^2}{(b_4-b_5)^2}  P^2 +
               \frac{b_4\omega_2-b_5\omega_1}{b_4-b_5} P - {\cal E} \ .   \label{Del4}
\end{eqnarray}
From (\ref{K1e}), we see that the first case occurs when the repeated root is at $K_1=0$. 

It should be emphasized that the occurence of a double root on the RHS of (\ref{K1e}) is a {\em necessary} 
condition for a bifurcation of the solution (giving rise to a transition between types) 
but not a {\em sufficient} condition.  For example, if the solution
is describing an oscillation on the interval between two adjacent roots of the RHS, and the two other
roots outside this interval merge, this will have no effect on the solution. We illustrate, in Figure 4,
with two concrete examples of equation (\ref{K1e}) emerging from the CQ system described in Section 2.
In both cases $Qa_0^{-2}=0.077$ and $ga_0^4=0$;
in the first case $r=1.01$ and in the second case $r=1.045$. In both cases we plot the roots of the RHS
as a function of the single remaining parameter $E$.  (The choice to plot the roots for fixed values of $Qa_0^{-2}$, $ga_0^4$
and $r$ and to vary $E$ is just an illustration; we could just as easilly vary any
of the other parameters or a combination thereof.)
In the first case  there are $4$ points $P_1,P_2,P_3,P_4$
at which there are double roots; however, transitions only occur at the two points $P_1,P_4$
(marked in Figure 4 with large dots). 
In the second case  there are $5$ points $P_1,P_2,P_3,P_4,P_5$
at which there are double roots; however, transitions only occur at the two points $P_2,P_4$. 
In both cases, the first transition is from type II to type I, and the second transition is from type I to type II,
as indicated by Roman numerals on the plot. 

The theoretical explanation of this is as follows.
In the first case, $r=1.01$, there are $4$ values of $E$ for which there is a double root. The points
labelled $P_3$ and $P_4$ on the diagram
are associated with the vanishing of the discriminant $\Delta_2$; the point labelled $P_1$ is a double root at $0$,
associated with the condition $\Delta_3=0$, and the point labelled $P_2$ is associated with the vanishing of the
discriminant $\Delta_1$. The motion takes place between the root that is at
$K_1\approx 0.00015$ and an adjacent root: for values of $E$ below $P_2$ the adjacent root is below,
for values of $E$ above $P_2$ the adjacent root is above. Thus the double root at $P_1$ indicates a value of $E$
for which there is a bifurcation, and the period of oscillation diverges.
The double root at $P_2$ is a special solution for which
$K_1$ and $K_2$ are constant (looking at (\ref{fe3})-(\ref{fe4}) it can be seen that there are $3$ kinds of solution
of this type, each corresponding to vanishing of one of the three factors on the RHS of this equation; these are related to the
{\em nonlinear normal modes} of the system \cite{mikhbar,pm1,pm2}). The point $P_2$  does
not, however, give rise to a transition in behavior of the CQ system;  the beating period diverges there, but the
type does not change. The double root at $P_3$ also does not mark a transition. This is precisely the case described above,
in which the oscillation is on the interval between 2 roots, and the other two roots outside this interval merge.
The point $P_4$, however, does mark a second transition, from type I beating back to type II. 

\begin{figure}[t]
\center{
\includegraphics[width=0.98\textwidth]{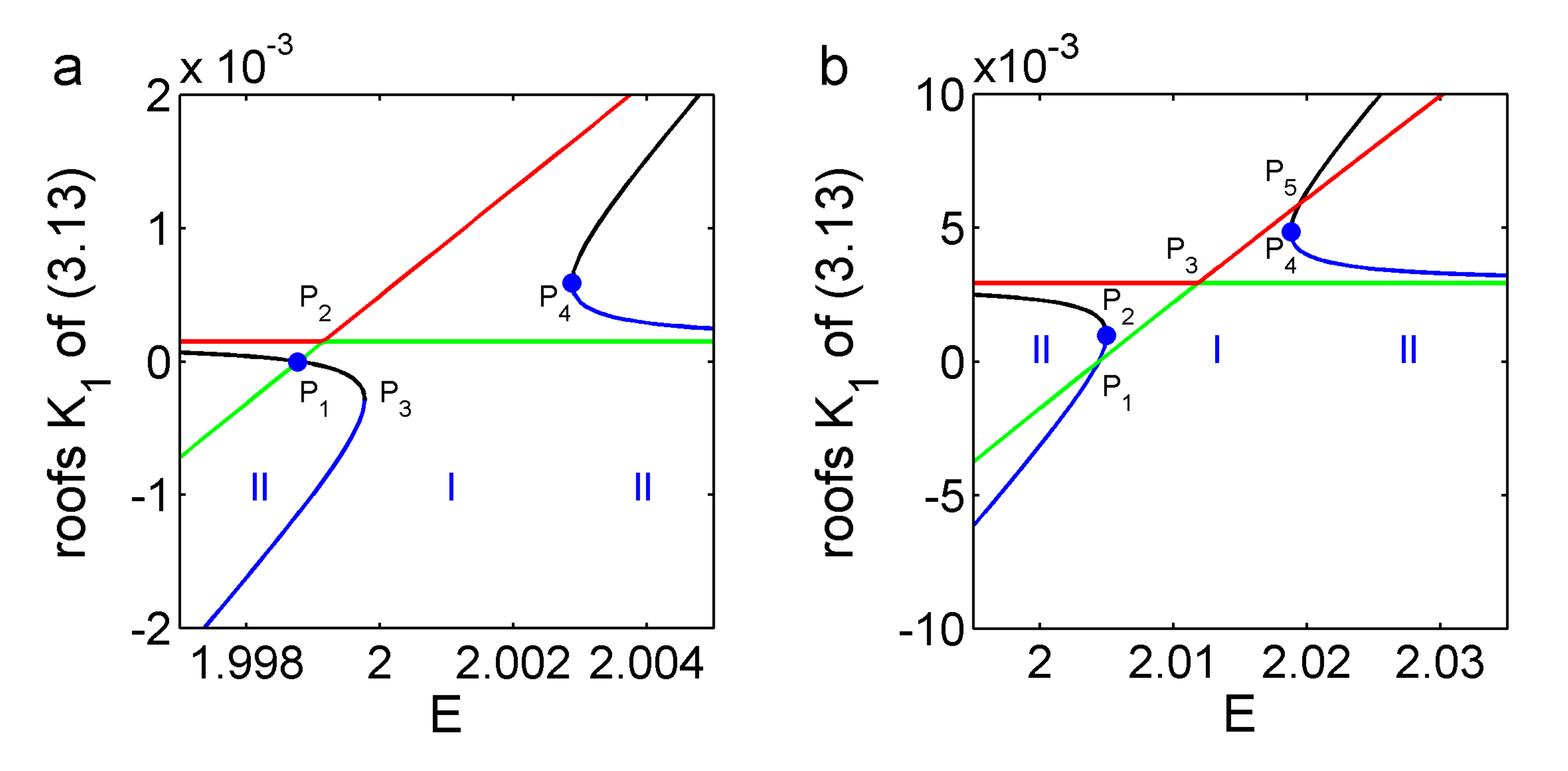}
}
\caption{\label{roots}
  Roots of the RHS of (\ref{K1e}) as a function of $E$ for the case of (\ref{K1e})
  emerging from the CQ system derived in Section 2, 
  with $Qa_0^{-2}=0.077$, $ga_0^4=0$ and
  (a) $r=1.01$, (b) $r=1.045$. All double roots are labelled, but double roots leading to a transition are marked
  with large (blue) dots. In both plots,
  red and green curves indicate roots of the first quadratic factor in (\ref{K1e}), and blue and black curves indicate roots
  of the second quadratic factor.   
  }
\end{figure}

Proceeding to the second example in Figure 4, there are now $5$ cases of a double root. $P_2$ and $P_4$ are
associated with the vanishing of the discriminant $\Delta_2$, $P_3$ with the vanishing of the discriminant $\Delta_1$.
$P_1$ is the case of a double root at zero associated with the condition $\Delta_3=0$, and $P_5$ is associated with
the final possibility, $\Delta_4=0$. There are however only $2$ transitions, associated with the points $P_2$ and $P_4$, for
similar reasons to the case described in the previous paragraph. 

In this section we have 
explained how small oscillations of the original Hamiltonian (\ref{H}) near its fixed point and 
near (symmetric) $1:1$ resonance can be approximated using the integrable Hamiltonian (\ref{H3})
and the single differential equation (\ref{K1e}). 
We have arrived at a simple analytic approximation for beating transitions, viz.  
{\em a necessary condition for  a transition between type I and
  type II beating is the vanishing of one of the four quantities $\Delta_1,\Delta_2,\Delta_3,\Delta_4$ given in
  (\ref{Del1}),(\ref{Del2}),(\ref{Del3}),(\ref{Del4}).} 
It should be emphasized that this far from trivializes the original problem.
There is substantial complexity hidden in the relationship between parameters 
and initial conditions of the original Hamiltonian and those of 
the integrable Hamiltonian. Also determining which of the vanishing conditions gives a
physical transition can be subtle. In  Section  4
we apply the approximation to the CQ and SAT models from Section 2
and validate its predictions against numerical results. 

\section{Application to the models}

\subsection{The CQ Model}

The potential of the CQ model, given   by (\ref{VCQ}), has an isolated minimum when 
$$ a_x = a_y = a_{\rm min} \equiv \frac{4 E\sqrt{2Q}}{3\sqrt{E-1}} C_0  $$
where $C_0>0$ is a solution of the equation
$$ 1 - C_0^2 =  \frac{1024E^4Q^2g}{81(E-1)^3} C_0^6  \ . $$
The $1:1$ resonance condition is $E= E_{\rm res}$ where
\begin{equation}
  E_{\rm res} = \frac{4}{1 + \sqrt{1 + \frac{65536 Q^2g}{81}}}  \label{rescon}
\end{equation}  
In the case of zero grade index, $g=0$, we have $C_0=1$ and the resonance condition is simply $E=2$. 
The model is valid if the parameters $E,Q,g$ are chosen so that $E\approx E_{\rm res}$ and 
the initial conditions (see (\ref{ic0})) satisfy $a_0\approx a_{\rm min}$ and $r\approx 1$.   

The relevant parameters for the quartic Hamiltonian (\ref{H1}) are 
\begin{eqnarray}  
  \omega_1^2 &=& \frac{81}{512}\ \frac{(E-1)^2((2-E)C_0^2+E-1)}{E^4Q^2C_0^6}  \nonumber\\
  \omega_2^2 &=& \frac{81}{512}\ \frac{(E-1)^3(3-2C_0^2)}{E^4Q^2C_0^6}  \nonumber\\
  a_1 &=&   \frac{243}{8192}\ 
  \frac{ (E-1) ^{5/2} \left( 2(E-3) C_0^{2}-3(E-1)  \right)}{ {Q}^{5/2}C_0^{7}{E}^{5} } \nonumber    \\
  a_2 &=&  \frac{243}{8192}\ 
  \frac { ( E-1) ^{7/2} (2C_0^{2}-5) }{ {Q}^{5/2}C_0^{7}{E}^{5}} \label{findas}  \\
  a_3 &=& \frac{729}{262144}\  
  \frac { (E-1) ^{3} ( 2(5-E)C_0^{2}+3(E-1) )  }{ C_0^{8}{E}^{6}{Q}^{3}}  \nonumber\\
  a_4 &=&  \frac {729}{131072}\
  \frac{(E-1)^3\left( 10(3-E)C_0^{2}+ 21(E-1) \right)  }{C_0^{8}{E}^{6} {Q}^{3} } \nonumber\\
  a_5 &=&  \frac {3645}{262144}\
  \frac{(E-1)^4( 7-2 C_0^{2} )}{C_0^{8}{E}^{6}{Q}^{3}}  \nonumber
\end{eqnarray}  

The detailed recipe for checking whether a given set of parameters and initial conditions $E,Q,g,a_0,r$ might give
rise to a transition is as follows:
\begin{enumerate}
\item  Compute the coefficients $\omega_1^2,\omega_2^2,a_1,a_2,a_3,a_4,a_5$ using  (\ref{findas}). This is the only
       stage of the recipe that is model dependent. 
       Compute the coefficients $b_1,b_2,b_3,b_4,b_5$ from  (\ref{bs}). 
\item  Compute the initial conditions $J_1(0),J_2(0)$ from (\ref{Jic}) and $\theta_1(0),\theta_2(0)$ from the comments
  following (\ref{Jic}). In the case of CQ, all the parameter values which we used gave $\theta_1(0)=0$ (we
  took $r>1$ throughout) and $\theta_2(0)=\pi$. 
\item  Compute the initial conditions $K_1(0),K_2(0)$ using (\ref{Keq1})-(\ref{Keq2}).   This is
  the only stage of the recipe that is not completely explicit, and involves solving two equations in two
  variables. If no real solution can be found, the method fails. A suitable initial guess for the solution
  is $K_1(0)\approx J_1(0)$ and $K_2(0)\approx J_2(0)$. 
\item Determine the value of ${\cal E}$, the constant value of the Hamiltonian $H_3$ using (\ref{H3}), taking
  $\cos(2(\phi_1-\phi_2))=1$. Determine the value of $P=K_1+K_2$.   
\item Compute $\Delta_1,\Delta_2,\Delta_3,\Delta_4$ from   (\ref{Del1}),(\ref{Del2}),(\ref{Del3}),(\ref{Del4}).
  Values of $E,Q,g,a_0,r$ for which any of these quantities vanish are candidates for transitions. 
\end{enumerate}  


Figure 5 displays results. Figure 5a shows numeric values and candidate analytic approximations of
$r_c$ as a function of $E$ for $Qa_0^{-2}=0.077$ and $ga_0^4=0.01$. The dots denote numeric values of transitions
in the original system.  The solid curves show candidate analytic approximations of 3 distinct types: 
(1) (black) values for which $\Delta_2=0$ (a closed loop with a cusp on the axis at $r=1$), 
(2) (green) values for which $\Delta_3=0$ (a simple open curve) and 
(3) (red) values for which $\Delta_4=0$ (two crossing open curves). For the values of
$Qa_0^{-2}$ and $ga_0^4$ specified, it seems there are two branches of parameter values for which
there are transitions. We denote
the lower branch (on the plot) by $r_{c,1}(E)$, which exists for
$E$ greater than a certain value which we denote by $E_{c,1}$, 
and 
the upper branch by $r_{c,2}(E)$, which exists for
$E$ greater than a certain value which we denote by $E_{c,2}$,
with $E_{c,2} \approx 1.975 < E_{c,1} \approx 1.977$. 
On the lower branch, as $E$ increases from $E_{c,1}$, $r_{c,1}(E)$ at first follows the approximation
$\Delta_2=0$, until a triple point at which the curves $\Delta_2=0$ and $\Delta_4=0$ intersect.
As $E$ increases further, $r_{c,1}(E)$ follows the approximation $\Delta_4=0$. Surprisingly, this
approximation stays reasonably accurate for the full range shown on the figure, even though
$r_{c,1}(E)$ rises to approximately $1.12$.
On the upper branch, as $E$ increases from $E_{c,2}$, $r_{c,2}(E)$ at first follows the approximation
$\Delta_3=0$, until a triple point at which the curves $\Delta_2=0$ and $\Delta_3=0$ intersect.
As $E$ increases further, $r_{c,2}(E)$ follows the approximation $\Delta_2=0$. However, the quality of this
approximation rapidly decreases as $E$ and $r_{c,2}(E)$ increase further, with the discrepancy already visible
on the plot for $r_c\approx 1.06$. 

\begin{figure}[ht]
\center{
\includegraphics[width=0.98\textwidth]{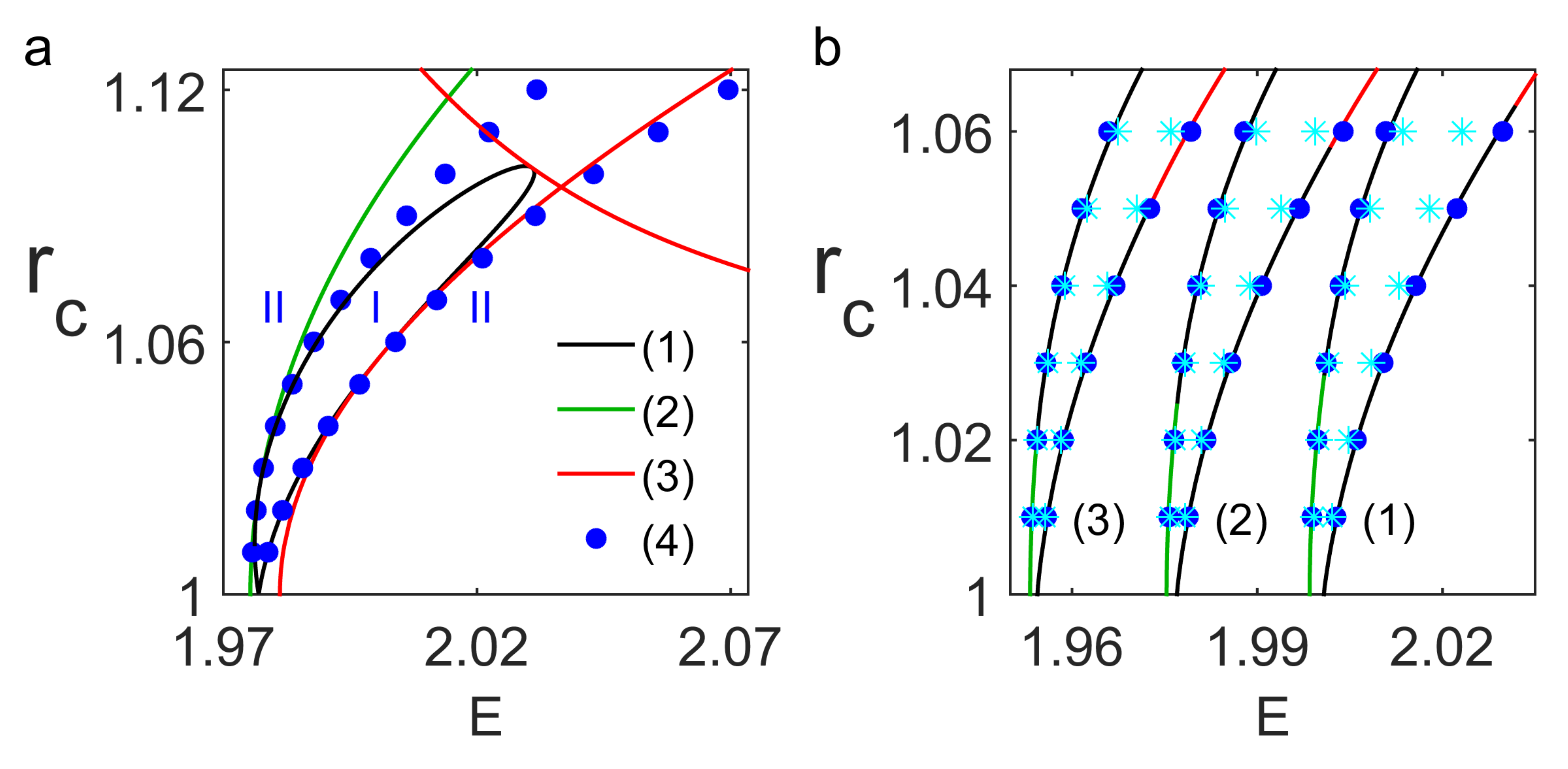}
}
\caption{\label{pv} Comparison of numerics and analytic approximation for the CQ system. (a) (Blue) dots denote numerical
  results and curves denote candidate analytic approximations for 
  $r_c$ as a function of $E$ for $Qa_0^{-2}=0.077$ and $ga_0^4=0.01$.
  Curve (1) (black) is $\Delta_2=0$, curve (2) (green) is $\Delta_3=0$, curve (3) (red) is $\Delta_4=0$.  
  The correct analytic approximation is made up of pieces of all the curves, see text for full details. 
  (b)  (Blue) dots denote numerical results for the exact Hamiltonian (\ref{H}), (turquoise) stars denote
  numerical results for the quartic Hamiltonian (\ref{H1}) and curves denote the analytic approximation for
  transitions, made up of pieces of the curves
  $\Delta_2=0$ (black), $\Delta_3=0$ (green),  $\Delta_4=0$ (red). $Qa_0^{-2} =0.077$ throughout.
  (1) $ga_0^4 = 0$, (2) $ga_0^4 = 0.01$, (3) $ga_0^4 = 0.02$.  
}
\end{figure}

Figure 5b shows numeric values and the {\em correct} analytic approximation (made up of pieces of the
curves $\Delta_2=0$, $\Delta_3=0$ and  $\Delta_4=0$) in the cases (1) $ga_0^4 = 0$, (2) $ga_0^4 = 0.01$, (3) $ga_0^4 = 0.02$,
all for $Qa_0^{-2}=0.077$. In addition, stars indicate numerical values of transitions obtained for the
quartic system with Hamiltonian (\ref{H1}). For small values of $r$, the numerical values for transitions
for the exact Hamiltonian and the approximate quartic Hamiltonian (\ref{H1}) are, as we would expect,
very close. However as $r$ increases, we see that the results for the quartic Hamiltonian rapidly
diverge from the results for the exact Hamiltonian, while, remarkably,  the analytic approximation continues
to be a reasonable approximation for the exact Hamiltonian. This may find an explanation in the fact that while the
exact Hamitlonian (\ref{H}), the quartic approximation (\ref{H1}) and the integrable approximation (\ref{H3}) all
agree close to the fixed point, the global properties of the exact Hamiltonian are expected to be closer to those of
the integrable approximation than the quartic approximation. 

Note also in Figure 5b that the intercepts of the curves on the $E$ axis, that we have denoted above by
$E_{c,1}$ and $E_{c,2}$, are very close to the values of $E$
determined by the resonance condition (\ref{rescon}), which are (1) $E=2$, (2) $E \approx 1.977$, (3)
$E\approx 1.954$.
However, even though the intercepts for the two curves obtained for each set of parameter
values are very close, they are not identical. This is something that is difficult to establish {\em a priori} by direct
numerics for the original systems (as the beating periods, for values of $r$ close to $1$, are very long), but once 
the analytic approximation is available to give accurate candidate values for the transition locations, it is
possible to verify them {\em a posteriori}. Thus 
{\em in the small band of values $E_{c,2}<E<E_{c,1}$ there is only a single beating transition as the beam eccentricity
is increased.} 
As $r$ is increased from $1$ there is immediately type I beating, and as $r$ is increased further there is only a
single transition to type II (as opposed, for example, to the sitution in Figure 2, where as $r$ is changed from $1$ type II
beating is seen, and then there are two transitions). Using the analytic approximation it can be shown (see
Appendix A) that the points $E_{c,1}, E_{c,2}$ are determined by the conditions
\begin{equation}
  \omega_1 - \omega_2 + P(b_2-2b_3 \mp b_5) = 0 \label{Ecconds}
\end{equation}  
(minus for $E_{c,1}$, plus for $E_{c,2}$) for a solution with $r=1$. (The condition $r=1$  implies $J_1(0) = K_1(0)=0$,
and then equation (\ref{Keq2}) gives a single equation from which to determine $K_2(0)$ from $J_2(0) = 
\omega_2 \left(a_0 - a_{\rm min} \right)^2$.)  
Figure 6 shows the dependence of $E_{c,1}$ and $E_{c,2}$
on $ga_0^4$ for two values of $Qa_0^{-2}$, as computed by the analytic approximation, along with a few numeric
values (computed {\em a posteriori}).
In addition the value of $E_{\rm res}$ from (\ref{rescon}) is shown, this being the value of $E$ for which there is 
exact $1:1$ resonance in the linear approximation.  
We see that the values of $E_{c,1}$, $E_{c,2}$ and $E_{\rm res}$ all decrease monotonically with $ga_0^4$. 

\begin{figure}[t]
\center{
\includegraphics[width=0.98\textwidth]{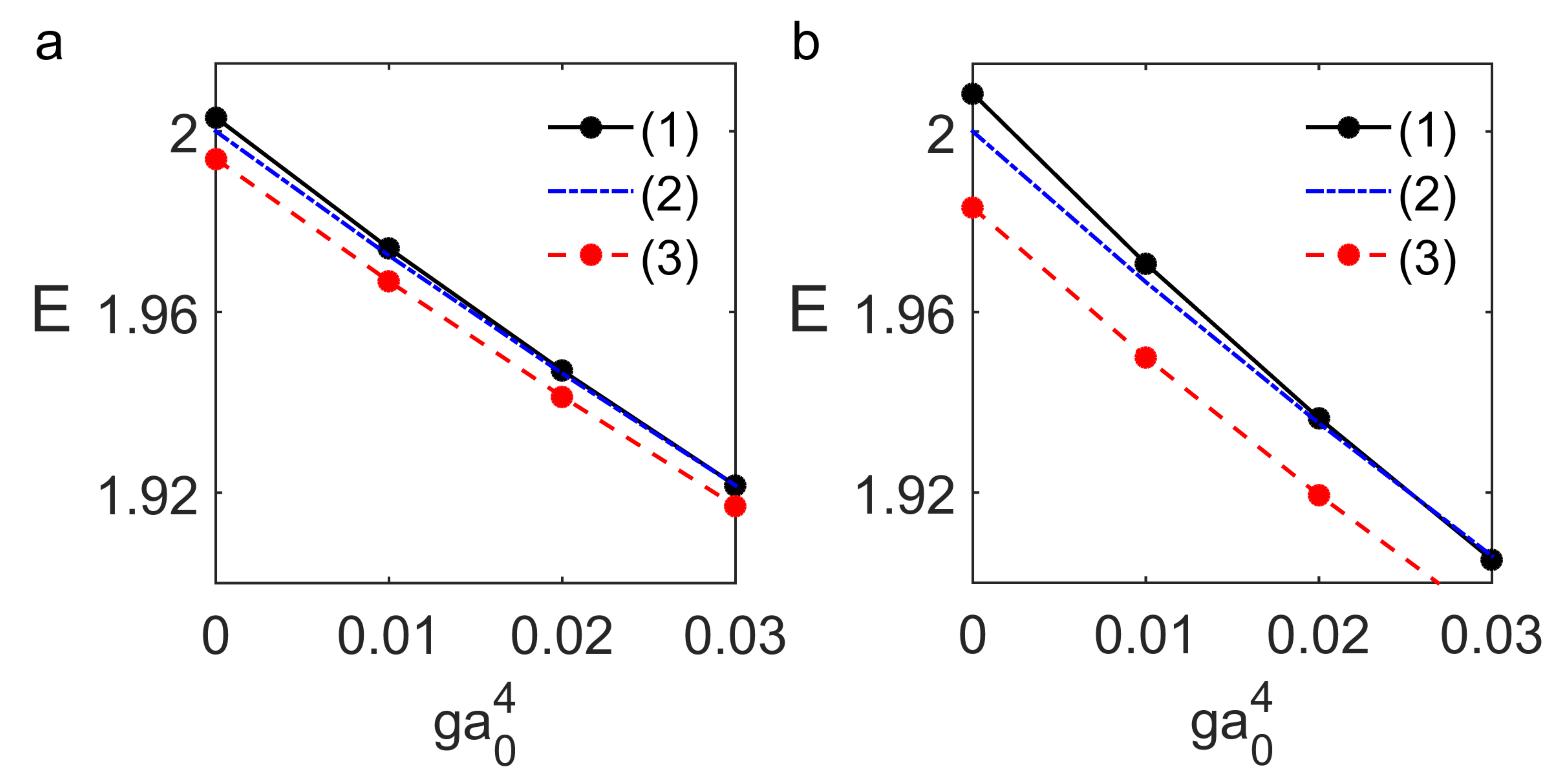}
}
\caption{\label{critEs2}
Dependence of $E_{c,1}$ and $E_{c,2}$ on $Qa_0^{-2}$ and $ga_0^4$, and comparison with 
the $E_{\rm res}$ from  (\ref{rescon}). (a) $Qa_0^{-2}=0.0836$ and (b) $Qa_0^{-2}=0.0924$.
In both plots
(1) (black, upper) shows  numeric and analytic values of  $E_{c,1}$,
(2) (blue, middle) shows $E_{\rm res}$, and 
(3) (red, lower) shows  numeric and analytic values of  $E_{c,2}$. 
}
\end{figure}

\subsection{The SAT Model}

The potential of the SAT model, given   by (\ref{VSAT}), has an isolated minimum when 
$$ a_x = a_y = a_{\rm min} \equiv   a_0 \sqrt{\frac{K_{\rm SAT}}{K_0}}  $$   
where $K_0$, which depends on the parameters $E,K_{\rm SAT}, ga_0^4$,  is a solution of the equation
\begin{equation}
\frac{K_0^2}{4E} + {\rm Li}_2 ( -K_0)  + \ln(1+K_0) -  \frac{K_{\rm SAT}^2ga_0^4}{4E}
= 0  \ .  \label{Hdef}
\end{equation}  
(Recall that the constant $K_{\rm SAT}$ is defined by $K_{\rm SAT}=4\alpha^2Ea_0^{-2}$.)
The resonance condition can be written $K_0 = K_{\rm res}$ where $K_{\rm res}$ is the solution of 
\begin{equation}  
   {\rm Li}_2 ( -K_{\rm res})  + 2\ln(1+K_{\rm res}) -  \frac{K_{\rm res}}{1+K_{\rm res}} = 0 \ .   
\end{equation}  
$K_{\rm res}$ has numerical value approximately $5.017$. 
We recall that for our analytic model to be most effective we need to be near resonance,
and the initial conditions should 
be close to the minimum, i.e. $a_0\approx a_{\rm min}$, or $K_{\rm SAT}\approx K_0$,  and $r\approx 1$.
These conditions give  $K_{\rm SAT} \approx K_{\rm res} = 5.017$  and $E \approx 6.550( 1 - ga_0^4)$. In practice
we will look at a large range of values of $E$ and $K_{\rm SAT}$, but focus on this
region. We also recall that in our model the sign of $\zeta_2(0)$ (as given in (\ref{ic1}) plays a critical role. 
From (\ref{Hdef}) we have $\zeta_2(0)=0$ (or equivalently $K_{\rm SAT}=K_0$) when  \cite{Ianetz2013PRA}
\begin{equation}
E =  \frac{ -K_{\rm SAT}^2(1-ga_0^4)} { 4( {\rm Li}_2(-K_{\rm SAT})+\ln(1+K_{\rm SAT})) }  \ .  \label{z20sign}
\end{equation}

The relevant parameters for the quartic Hamiltonian (\ref{H1}) in the SAT case are 
\begin{eqnarray}  
  \omega_1^2 &=& \frac{2}{a_0^4}\left( \frac{K_0^2}{K_{\rm SAT}^2} + ga_0^4 \right)  
  \nonumber\\
  \omega_2^2 &=& \frac{4}{a_0^4K_{\rm SAT}^2}\left(  -2E\ln(1+K_0) + K_0^2  + \frac{2  E K_0}{1+K_0} \right)  \nonumber\\
  a_1 &=&  \frac{\sqrt{2K_0}}{K_{\rm SAT}^{5/2}a_0^5}  \left( 2E\ln(1+K_0) - 3K_0^2-\frac{2EK_0}{1+K_0} \right)  \nonumber\\
  a_2 &=&  \frac{\sqrt{2K_0}}{3K_{\rm SAT}^{5/2}a_0^5}  \left(-2E\ln(1+K_0)  -3K_0^2  + \frac{2EK_0(1+3K_0)}{(1+K_0)^2}   \right) 
  \label{asforSAT}  \\
  a_3 &=&  \frac{K_0}{4K_{\rm SAT}^3a_0^6} \left( -2E\ln(1+K_0) + 5K_0^2 +  \frac{2EK_0}{1+K_0}  \right)  
     \nonumber\\
  a_4 &=&  \frac{K_0}{2K_{\rm SAT}^3a_0^6} \left( -2E\ln(1+K_0) + 15K_0^2 + \frac{2EK_0(1-K_0)}{(1+K_0)^2}  \right)  
     \nonumber\\
  a_5 &=&  \frac{K_0}{12K_{\rm SAT}^3a_0^6} \left( 2E\ln(1+K_0) + 15K_0^2 -  \frac{2EK_0(1+10K_0 + K_0^2)}{(1+K_0)^3}  \right)  \ . 
    \nonumber 
\end{eqnarray}  

\begin{figure}[t]
\center{
\includegraphics[width=0.98\textwidth]{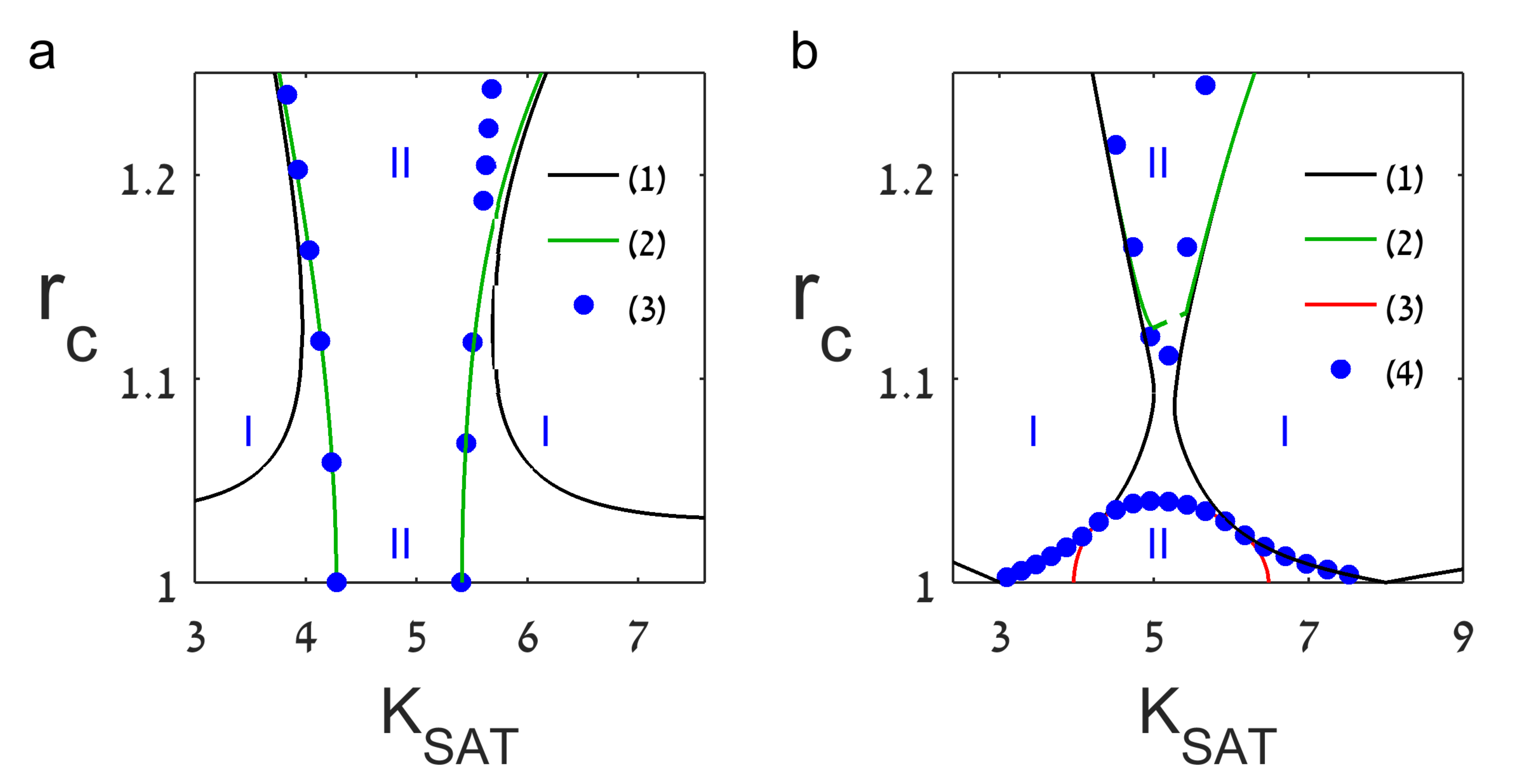}
}
\caption{\label{oldfig6}
  Behavior of $r_c$ as a function of $K_{\rm SAT}$ for the SAT model for fixed values of $E$ and $ga_0^4$,
  comparison of numerics and analytics. (a) (left) $E=6.3$, $ga_0^4=0$ (below the critical threshold). 
  (b) (right) $E=6.7$, $ga_0^4=0$ (above the critical threshold). In both plots (blue) dots indicate
  transtions obtained from numerics of the original system. Curves denote different degeneracies in the
  analytic model:  (1) (black) $\Delta_2=0$,  (2) (green) $\Delta_3=0$, and  (3) (red) $\Delta_4 = 0$ (for plot (b) only). 
}
\end{figure}

The method is identical to that given for CQ in the previous subsection, so we can immediately present 
results. For fixed values of $E$ and $ga_0^4$ we look for values of $r$ giving
beating transitions as a function of $K_{\rm SAT}$. Both numeric and analytic results 
suggest there is a qualitative difference in behavior for $E$ above and below a critical threshold, 
and our results are consitent with the value of this threshold being  approximately $6.550( 1 - ga_0^4)$,
as found above. Figure 7 displays results for $ga_0^4=0$ and $E=6.3$ (below the threshold, left) and
$E=6.7$ (above the threshold, right). The numeric results show that below the threshold, there are two
ranges of $K_{\rm SAT}$ for which there is a single beating transition, from type I (for $r$ below $r_c$)
to type II (for $r$ above $r_c$). For values of $K_{\rm SAT}$ below or above these two ranges, there is only type I
beating, and for values between the two ranges there is only type II beating.  The analytic approximation
reproduces these results well. In this region of parameter space there are values for which $\Delta_2=0$
(indicated in black in the figure) and
$\Delta_3=0$ (indicated in green). It is the latter that are physically relevant, and the
values of $r_c$ predicted by the analytic model are accurate for a good range. 
Moving ``above the threshold'', numerics show there is a range of values of $K_{\rm SAT}$ for which, as
$r$ is increased from $1$, the beating is initially type II, then there is a transition to type I. For some of these
values there is then a further transition back to type II for quite high values of $r$. It should be mentioned
that these latter transitions were initially discovered using the analytic approximation, and 
confirmed numerically {\em a posteriori}. The analytic approximation reproduces the first transition very well, using pieces of
the $\Delta_2=0$ and $\Delta_4=0$  degeneracy curves. The upper transition is not reproduced well, which is
not surprising bearing in mind the values of $r$ involved. Pieces of the
$\Delta_2=0$ and $\Delta_3=0$  degeneracy curves are close to some of the results, but for a small
range of values of $K_{\rm SAT}$ and $r$ the model fails
as there is no solution of equations (\ref{Keq1})-(\ref{Keq2}). Two branches of the $\Delta_3=0$ degeneracy
curve come to an abrupt end
(in the plot we have connected the ends with a dashed line, which is not associated
with any degeneracy).
The values of parameters involved
are precisely those for which $\zeta_2(0)\approx 0$.

\begin{figure}[t]
\center{
\includegraphics[width=0.8\textwidth]{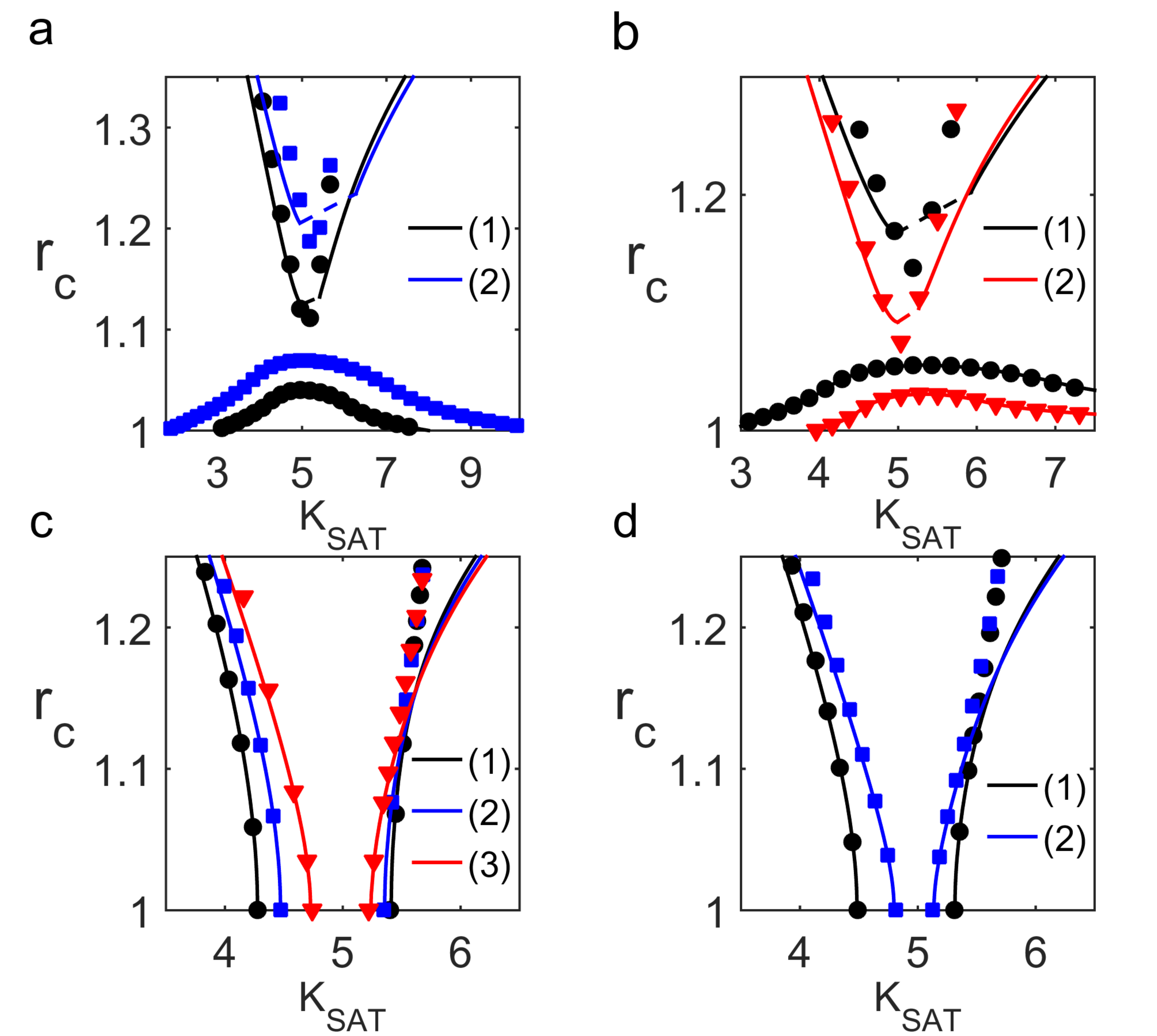}
}
\caption{\label{figure8}
Behavior of $r_c$ as a function of parameters $K_{\rm SAT}$, $E$ and $ga_0^4$  for the SAT model, 
comparison of analytic approximation (curves) and numerics (blue dots).  
(a) $ga_0^4=0$, (1) (black) $E=6.7$, (2) (blue) $E=7.0$.   
(b) $ga_0^4=0.02$, (1) (black) $E=6.7$, (2) (red) $E=6.5$.
(c) $ga_0^4=0$, (1) (black) $E=6.3$, (2) (blue) $E=6.4$, (3) (red) $E=6.5$.
(d) $ga_0^4=0.02$, (1) (black) $E=6.3$, (2) (blue) $E=6.4$.  
}
\end{figure}

Figure 8 enlarges upon these results for different values of $E$ and $ga_0^4$. In the 4 panels here, the upper panels
(a and b) show results for values of $E$ above the threshold, and the lower panels (c and d) show results for values
of $E$ below the threshold. In the left panels (a and c), $ga_0^4=0$, in the right panels (b and d), $ga_0^4=0.02$.
For
$ga_0^4=0$, the values $E=6.3,6.4,6.5,6.7,7.0$ are shown, the first three of which are below the threshold (in panel c),
and the last two above the threshhold (in panel a). 
For
$ga_0^4=0.02$, the values $E=6.3,6.4,6.5,6.7$ are shown, the first two of which are below the threshold (in panel d),
and the last two above the threshold (in panel b). 
Note specifically that for $ga_0^4=0$ the case $E=6.5$ is below the threshold (approximately $6.55$), while for
$ga_0^4=0.02$ it is above (as the threshold drops to approximately $6.42$). Thus (for example) for $E=6.5$, $K_{\rm SAT}=5$ and
$ga_0^4=0$, no beating transitions are observed as the beam eccentricity is increased; but if the grade index is
changed to $ga_0^4=0.02$, there are two beating transitions.
The analytic theory fully explains this phenomenon. Indeed, for all the cases shown in Figure 8,
the analytic theory  is in excellent quantitative
agreement 
with numerics for lower values of $r$, and gives reasonable qualitative predictions 
for higher values of $r$. 

Another conclusion from Figure 8 is that for values of  $E$ below the threshold, we can find two
values of $K_{\rm SAT}$ that give rise to a given value of $r_c$, but for $E$ above the threshold this need not
be the case; furthermore the gap in $r_c$ values increases with the given value of $E$. In Figure 9
we illustrate this phenomenon more clearly. For the case $ga_0^4=0$,
we show contours in the $K_{\rm SAT},E$ plane that give rises to the values $r_c = \frac1{0.95}\approx 1.053$ (black),
$r_c = \frac1{0.92}\approx 1.087$ (blue) and   $r_c = \frac1{0.895}\approx 1.117$ (red). It is clear that the ``gap'' 
between the two branches of each contour increases with $r$.  Note that in the upper branch of each contour there is a
small section denoted by a dashed line where the analytic method fails
(the dashed line is a straight line between
the last two points on each side for which the method works). As expected,
the regions where the method  fails straddle the curve (\ref{z20sign}),
incidicated by a dashed turquoise curve.  Note further that in many cases the analytic method works well far beyond
the region in which this is expected, but there are some exceptions. 

\begin{figure}[t!]
\center{
\includegraphics[width=0.7\textwidth]{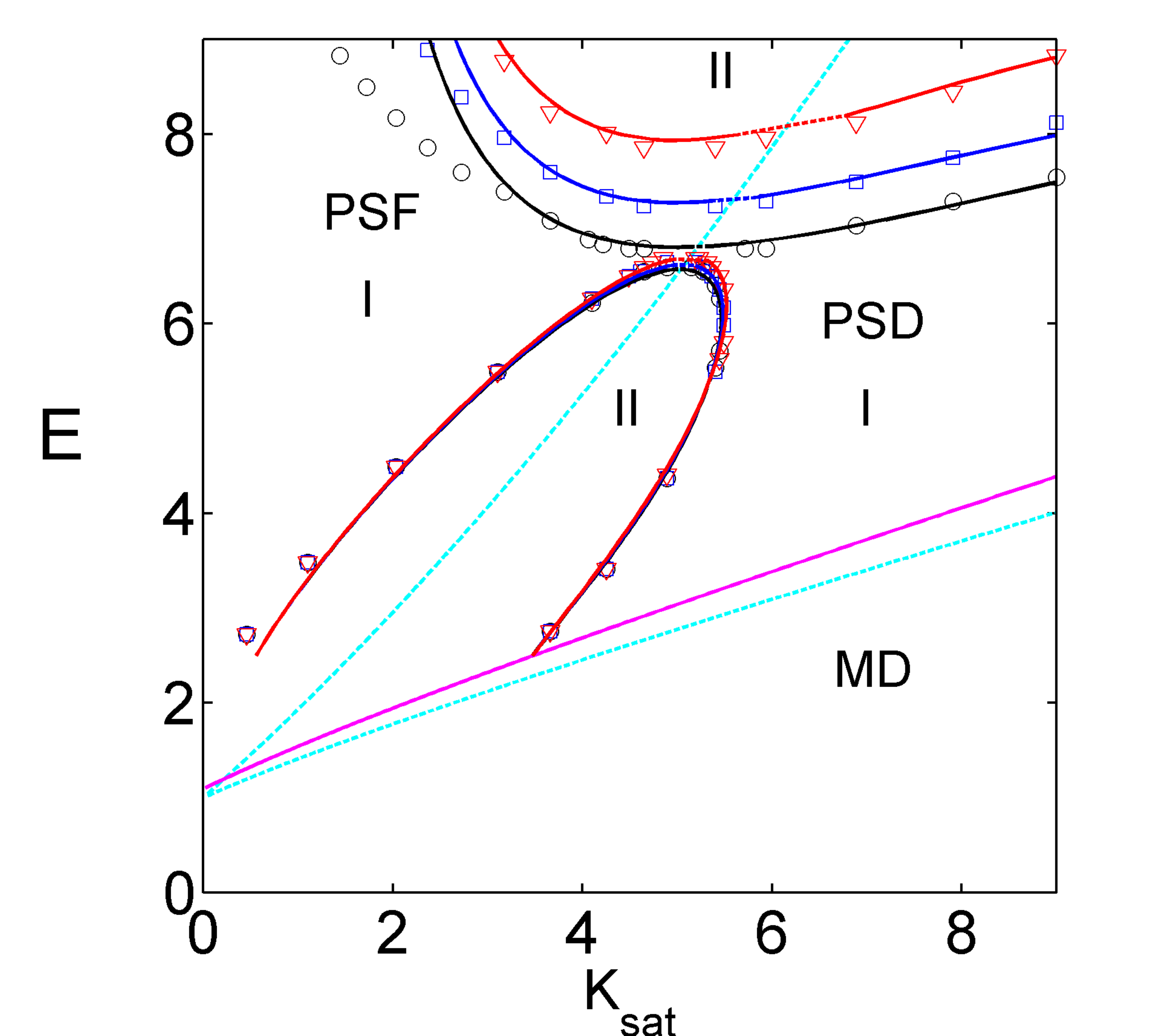}
}
\caption{\label{figure9}
  Contours of constant $r_c$ in the  $K_{\rm SAT},E$ plane for the SAT model with $ga_0^4=0$, 
  comparison of analytic approximation (curves) and numerics (data points).  
  Red $r_c=\frac1{0.895}\approx1.117$, blue $r=\frac1{0.92}\approx 1.087$, black $r=\frac1{0.95}\approx 1.053$.
  The upper dashed turquoise curve is the curve (\ref{z20sign}). In the case of a symmetric beam, this curve
  divides between periodic self-focusing (PSF) and periodic self-diffusing (PSD) solutions, see
  \cite{Ianetz2010PRA_2}. The lower dashed
  turquoise curve divides between PSD and monotonic diffracting (MD) solutions (in the case
  of symmetric beams; as asymmetry is introduced the curves moves). 
}
\end{figure}

\section{Conclusions and discussion} 
In this paper we have described the beating phenomena observed in the equations of motion for the beam
widths obtained in a collective variable approximation to solution of the GNLSEs relevant for beams 
in nonlinear waveguides  with cubic-quintic (CQ) and saturable (SAT) nonlinearities and a graded-index profile.
We have described the different types of beating, and the transitions between them.
Arguing that the origin of these phenomena is in a $1:1$ Hamiltonian resonance, we have developed an
approximation scheme for small oscillations in 
a class of 2 degree-of-freedom Hamiltonian systems with an isolated fixed point close to $1:1$ resonance.
We have shown that such oscillations can be described by an integrable Hamiltonian, or, alternatively, a single
first order differential equation (\ref{K1e}). Understanding the bifurcations of the system, which include the
beating transitions, can be reduced to looking at the bifurcations of the roots of a pair of quadratic equations. 
Applying our general methodology to the specific cases of the CQ and SAT models we managed to reproduce
numerical results for beating transitions over a large range of parameter values. The theory
allows us to map out the regions (of parameter space and beam eccentricities) 
where beating transitions do and do not exist. Amongst other things,
in the CQ case we identified a band of beam energies
for which there is only a single beating transition (as opposed to $0$ or $2$) as the beam eccentricity is increased,
and  in the SAT case we explained the appearance and disappearance of transitions with
changes of the grade-index. 

We expect our methods to have applications to related problems in nonlinear optics, for
nonlinearities other than the ones studied here,  for different beams, such as super-Gaussian
beams \cite{supergauss}, and for optical bullets \cite{Kivshar2003book,bullets}. We are encouraged by the fact
that there is some recent experimental evidence
\cite{skarkarobust} of breathing in optical solitons, albeit in a dissipative setting.
We also hope the general theory of $1:1$ resonances that we have developed will find application
in the settings of nonlinear mechanics and astronomy, as well as suitable extensions for 
$1:1:1$ resonances in higher dimensional systems (see for example the recent papers \cite{JS1,JS2}). 

\newpage

\appendix 

\section{Further Technical Details}

As explained in section 3, the Hamiltonian (\ref{H3}) is in an integrable approximation
to the Hamiltonian (\ref{H2}), and is obtained from (\ref{H2}) via a canonical transformation
and neglecting higher order terms. The only need for explicit details of the canonical transformation is
to compute the initial conditions of the variables $K_1,K_2$ from the initial conditions 
of $J_1,J_2$ given in (\ref{Jic}). The equations to be solved are 
\begin{eqnarray}
  J_1 &=&  K_{{1}}  
  \mp  {\frac {4K_{{1}}a_{{1}}}{4\omega_{{1}}^{2}- \omega_{{2}}^{2}}\sqrt {{\frac {2K_{{2}}}{\omega_{{2}}}}}}  
    + \left(   
\frac{a_1^2 \left( 48\omega_1^{4}-8\omega_1^{3}\omega_2-40\omega_2^{2}\omega_1^{2}+2\omega_1\omega_2^{3}+5\omega_2^{4} \right)}
    {4\omega_{{2}}^{2}\omega_{{1}}^{3} \left( 2\omega_{{1}}+ \omega_{ {2}} \right)^{2} \left( 2\omega_{{1}}-\omega_{{2}}\right) ^{2}} 
    -{\frac {5a_{{3}}}{{2\omega_{{1}}}^{3}}}
    \right)  K_1^2  \nonumber \\
&&  + \left(  \frac { \left( 40\omega_{{1}}^{3}+28\omega_{{1}}^{2}\omega_{{2}}-6\omega_{{1}}\omega_{{2}}^{2}
        -3\omega_{{2}}^{3} \right) a_{{1}}^{2} }
          {\omega_{{1}}^{2}\omega_{{2}} \left( 2\omega_{{1}}-\omega_{{2}} \right) ^{2} \left( 2\omega_{{1}}+\omega_{{2}} \right)^{2}
                  \left( \omega_{{1}}+\omega_{{2}} \right) }
\right.  \nonumber \\
&& \left. 
+\frac { \left( 12\omega_{{1}}^{3}+11\omega_{{1}}^{2}\omega_{{2}}-10\omega_{{1}}\omega_{{2}}^{2}
             -6\omega_{{2}}^{3} \right) a_ {{1}}a_{{2}}} 
{2\omega_{{1}}^{2}\omega_{{2}}^{3} \left( \omega_{{1}}+\omega_{{2}} \right)  \left( 4\omega_{{1}}^{2}-\omega_{{2}}^{2}\right) }
- \frac {\left( 3\omega_{{1}}+2\omega_{{2}} \right) a_{{4}}}
        {2\omega_{{1}}^{2}\omega_{{2}} \left( \omega_{{1}}+\omega_{{2}} \right) }  \right) K_1K_2  \ ,  \label{Keq1}\\
 J_2 &=&  K_{{2}}  
 \mp 2 \left(
   \frac { \left( 2\omega_{{1}}^{2}-\omega_{{2}}^{2} \right)K_{{1}}a_{{1}}}
        {\omega_{{1}}\omega_{{2}} \left( 4\omega_{{ 1}}^{2}-\omega_{{2}}^{2} \right) }
+  \frac{K_{{2}}a_{{2}}}{\omega_{{2}}^{2}}
\right)  \sqrt{\frac {2K_{{2}}}{\omega_{{2}}}}
   + \left(  \frac{33a_2^2}{4\omega_2^5} -\frac {5a_5}{2\omega_{{2}}^{3}}  \right) K_2^2  \nonumber \\
&& 
   + \frac{\left( 16\omega_{{1}}^{4}+8\omega_{{1}}^{3}\omega_{{2}} - 12\omega_{{1}}^{2}\omega_{{2}}^{2}
               -2\omega_{{1}}\omega_{{2}}^{3}+3\omega_{{2}}^{4} \right) {a_{{1}}}^{2}{K_{{1}}}^{2}}
   {2{\omega_{{1}}}^{2}\omega_{{2}}^{3} \left( 2\omega_{{1}}-\omega_{{2}} \right)^{2}\left(2\omega_{{1}}+\omega_{{2}}\right)^{2}   }
      \nonumber\\ 
&&
+ \left(  
\frac {\left( 8\omega_{{1}}^{4}+16\omega_{{1}}^{3}\omega_{{2}}-10\omega_{{1}}^{2}\omega_{{2}}^{2}
               -8\omega_{{1}}\omega_{{2}}^{3}+\omega_{{2}}^{4} \right) {a_{{1}}}^{2}}
       {\omega_{{1}}^{2}\omega_{{2}}^{2} \left( 2\omega_{{1}}-\omega_{{2}} \right)^{2}
         \left( 2\omega_{{1}}+2\omega_{{2}} \right) ^{2} \left( \omega_{{1}}+ \omega_{{2}} \right)        }
\right. \nonumber\\
&& \left.       
+\frac { \left( 40\omega_{{1}}^{3}+44\omega_{{1}}^{2}\omega_{{2}}-9\omega_{{1}}\omega_{{2}}^{2}-16\omega_{{2}}^{3} \right) a_{{1}}a_{{2}}}
{2\omega_{{1}}\omega_{{2}}^{4} \left( \omega_{{1}}+\omega_{{2}} \right) 
\left( 2\omega_{{1}}-\omega_{{2}} \right)\left( 2\omega_{{1}}+\omega_{{2}} \right)  
} 
-\frac { \left( 2\omega_{{1}}+3\omega_{{2}} \right) a_{{4}}}
       {2\omega_{{1}}\omega_{{2}}^{2} \left( \omega_{{1}}+\omega_{{2}} \right) }
\right) K_1K_2 \ . \label{Keq2}
\end{eqnarray}  
Here the upper signs should be taken in the square roots terms 
in the case $\theta_2(0)=0$ and the lower signs in the case $\theta_2(0)=\pi$. 

In Section 4.1, in the study of the CQ system, we stated the conditions (\ref{Ecconds}) for the value $r_c$
giving a beating transition to tend to $1$. We briefly describe the origin of these conditions.
The symmetric
solutions with $a_x=a_y$ of (\ref{em}), arising from the initial condition $r=1$, 
correspond to solutions with $K_1\equiv 0$ of (\ref{fe1})-(\ref{fe4}). From (\ref{K1e}), the values of $P$ and
${\cal E}$ for such a solution must evidently satisfy $b_3P^2 + \omega_2 P -{\cal E}=0$, which is just the condition
$\Delta_3=0$, see (\ref{Del3}).  As explained in Section 3, a necessary condition for a beating transition is
the vanishing of one of the quantites $\Delta_1,\Delta_2,\Delta_3,\Delta_4$.  To determine $E_{c,1}$ in Section 4.1
we want $r_c\rightarrow 1$ for a solution of $\Delta_{2}=0$. Clearly this requires $\Delta_2=\Delta_3=0$, 
and some simple algebra then gives the condition $\omega_1-\omega_2 + P(b_2-2b_3-b_5)=0 $.  To determine $E_{c,2}$, 
however, is not so straightforward, as for this we want
we want $r_c\rightarrow 1$ for a solution of $\Delta_{3}=0$, and apparently we do not have two equations. The resolution
of this conundrum is as follows: Although we stated above that the symmetric solutions  of (\ref{em}) 
correspond to solutions with $K_1\equiv 0$ of (\ref{fe1})-(\ref{fe4}), the latter in fact provide a {\em blow up} of
the former --- there is a $3$ parameter family of the latter and only a $2$ parameter family of the former.
Solving (\ref{fe1})-(\ref{fe4}) in the case $K_1\equiv 0$, we obtain $K_2=P$ (constant),
$\phi_2 = \phi_{2}(0) + (\omega_2 + 2 b_3P)z  $, and that $\phi_1$ must satisfy the ODE
$$   \dot{\phi}_1 = \omega_1 + b_2 P + b_5 P \cos \left( 2\left(  \phi_{2}(0) + (\omega_2 + 2 b_3P)z  - \phi_1 \right)\right)  \ . $$
This latter equation can be solved explicitly, and for a general choice of the constant of integration
will give a complicated function $\phi_1(z)$. However, for a beating transition 
we seek a solution that is characterized by a single frequency, i.e. we need 
$$ \phi_1(z) =  \phi_1(0) +  (\omega_2 + 2 b_3 P)  z   $$
Substituting this in the differential equation, we obtain 
$$   \omega_2 + 2 b_3 P  =  \omega_1 + b_2 P + b_5 P \cos \left( 2\left(  \phi_{2}(0)  - \phi_1(0) \right)\right)  \ . $$
Since the initial conditions 
$\phi_1(0),\phi_2(0)$ take the values $0$ or $\pi$, we deduce that $\omega_1 - \omega_2 + P(b_2 - 2b_3 + b_5) = 0$,
as required. 

\section{A two time expansion approach}

In this appendix we outline a two time expansion approach \cite{kc,ManMan}
which is an alternative to the procedure  based on canonical transformations
given in Section 3.

We wish to look at solutions of the Hamiltonian system with Hamiltonian (\ref{H}) and initial conditions (\ref{ic0}).
We assume the system has an isolated symmetric minimum at which the system is close to $1:1$-resonance. To apply a
two time technique we need to introduce a small parameter $\epsilon$ explicitly into the equations. Our systems involve
a number of system parameters, for example in the CQ case, the parameters $E,Q,g$, for which the resonance condition is
(\ref{rescon}). We introduce a small parameter by selecting one system parameter and writing this
as its value at resonance plus a small perturbation. However, for reasons described in \cite{Verhulst1}, the ``small perturbation'' here
  should be {\em quadratic} in the small parameter. Thus, for example in CQ, we have to consider two possibilities,
$ E = E_{\rm res} \pm \epsilon^2 $
  where $E_{\rm res}$ (which depends on the other system parameters $Q,g$) is the value of $E$ at resonance. The two resulting
  expansions will differ just in signs. This is the counterpart in the two time method of the need to choose $\theta_2(0)$
  to be $0$ or $\pi$ in Section 3 and the resulting choice of signs in equations (\ref{Keq1})-(\ref{Keq2}). 
  However, we emphasize that it is not the same, so the resulting method is different, in particular, the ``choice'' in Section 3
  involves the initial conditions as well as the system parameters.

  Taking, as before,  the minimum of the potential $V$ to be at $a_x=a_y=a_{\rm min}$ we now write
  $$  a_x = a_{\rm min} + \epsilon \tilde{a}_x \ , \qquad a_y = a_{\rm min} + \epsilon \tilde{a}_y  $$
  and expand to 4th order in $\epsilon$. The order $0$ terms are irrelevant and can be discarded. 
  The order $1$ terms vanish by definition  of $a_{\rm min}$.
  In the other terms there is dependence on all the system parameters. However by making the
  assignment of the form $E = E_{\rm res} \pm \epsilon^2$, discarding all terms of order higher than $4$
  and a suitable rescaling, we obtain an approximate potential of the form 
\begin{eqnarray*}
  \tilde{V} &=& \frac12 C_1  (\tilde{a}_x^2 +   \tilde{a}_y^2)       
  + \epsilon  \left( C_2( \tilde{a}_x^3 +   \tilde{a}_y^3)  + C_3\tilde{a}_x\tilde{a}_y  (\tilde{a}_x +   \tilde{a}_y)   \right)  \\  
&&  + \epsilon^2 \left( C_4( \tilde{a}_x^4 +   \tilde{a}_y^4)  + C_5\tilde{a}_x\tilde{a}_y (\tilde{a}_x^2 +   \tilde{a}_y^2)
     + C_6 \tilde{a}_x^2\tilde{a}_y^2   +  C_7(\tilde{a}_x^2 +   \tilde{a}_y^2)       + C_8 \tilde{a}_x\tilde{a}_y   \right) \ .   
\end{eqnarray*}
Here $C_1,\ldots,C_8$ are all functions of the system parameters excluding the parameter replaced by $\epsilon$. 
Note that as a result of the dependence of the system parameters on $\epsilon$ there are now quadratic terms
in $\tilde{a}_x,\tilde{a}_y$ in the $O(\epsilon^2)$ terms. 

Following the usual two time formalism, we seek solutions of the system with potential $\tilde{V}$ in the form 
\begin{eqnarray*}
  \tilde{a}_x &=& \Lambda_1(\epsilon^2 z)  \cos (\sqrt{C_1}z) + \Lambda_2(\epsilon^2 z) \sin (\sqrt{C_1}z) +
       \epsilon \tilde{a}_{x,1} (z,\epsilon^2 z) + \epsilon^2 \tilde{a}_{x,2} (z,\epsilon^2 z) + \ldots \\ 
  \tilde{a}_y &=& \Lambda_3(\epsilon^2 z) \cos (\sqrt{C_1}z) + \Lambda_4(\epsilon^2 z) \sin (\sqrt{C_1}z) +
  \epsilon \tilde{a}_{y,1} (z,\epsilon^2 z) + \epsilon^2 \tilde{a}_{y,2} (z,\epsilon^2 z) + \ldots
\end{eqnarray*}
Here $\Lambda_1(\epsilon^2 z), \Lambda_2(\epsilon^2 z), \Lambda_3(\epsilon^2 z), \Lambda_4(\epsilon^2 z)$ are
functions of the slow variable $\epsilon^2 z$. Substituting in the equations of motion and equating order-by-order,
the first order terms $\tilde{a}_{x,1}, \tilde{a}_{y,1}$ can be determined, and a system of first order equations
is obtained that $\Lambda_1,\Lambda_2,\Lambda_3,\Lambda_4$ must satisfy to guarantee the absence of secular terms
in $\tilde{a}_{x,2}, \tilde{a}_{y,2}$. Writing
\begin{eqnarray*}
  R_1 &=& \Lambda_1^2 + \Lambda_2^2 + \Lambda_3^2 + \Lambda_4^2  \\
  R_2 &=& \Lambda_1^2 + \Lambda_2^2 - \Lambda_3^2 - \Lambda_4^2  \\
  R_3 &=& \Lambda_1 \Lambda_3  + \Lambda_2 \Lambda_4  \\
  R_4 &=& \Lambda_1 \Lambda_4  - \Lambda_2 \Lambda_3
\end{eqnarray*}
(c.f. \cite{RPV,pm2,hh}) we obtain the system
\begin{eqnarray}
  R_1' &=& 0  \nonumber  \\   
  R_2' &=& 4 R_4( \gamma_1 R_1 + \gamma_2 + (\gamma_3+\gamma_4)R_3) \label{Reqs} \\
  R_3' &=& -\gamma_3 R_2 R_4   \nonumber \\
  R_4' &=&   - R_2( \gamma_1 R_1 + \gamma_2 + \gamma_4 R_3)   \nonumber
\end{eqnarray}
where the constants $\gamma_1,\gamma_2,\gamma_3,\gamma_4$ are certain combinations of the constants $C_1,\ldots,C_8$. 
(Note $R_3^2 + R_4^2 = \frac14 ( R_1^2 - R_2^2 )$.)  Thus $R_1$ is an invariant, as are the quantities 
$$ Q_2 = R_2^2 + 4\left(1 + \frac{\gamma_4}{\gamma_3}  \right)
                  \left(  R_3 + \frac{\gamma_1 R_1 + \gamma_2}{\gamma_3+\gamma_4}  \right)^2 
\ ,
\qquad   Q_3  = R_4^2  -  \frac{\gamma_4}{\gamma_3}
\left( R_3 + \frac{\gamma_1 R_1+ \gamma_2}{\gamma_4}  \right)^2 \ . 
$$ 
Note that $R_1,Q_2,Q_3$ are related by
$$ Q_2 + 4Q_3 = R_1^2 - \frac{4(\gamma_1R_1+\gamma_2)^2}{\gamma_4(\gamma_3+\gamma_4)}\ .  $$ 
Using the invariants it is possible to write a single differential equation for the quantity $R_3$:
\begin{equation}
(R_3')^2 =    -4 \gamma_4 \left(\gamma_3 + \gamma_4  \right)
\left(   \left(  R_3 + \frac{\gamma_1 R_1 + \gamma_2}{\gamma_3+\gamma_4}  \right)^2
     -\frac{\gamma_3}{4(\gamma_3 + \gamma_4 )}    Q_2    \right) 
\left(  \left( R_3 + \frac{\gamma_1 R_1+ \gamma_2}{\gamma_4}  \right)^2
     + \frac{\gamma_3  Q_3}{\gamma_4}   \right)\ . 
\label{R3eq}\end{equation}
This has the same form as (\ref{K1e}) ---  the right hand side is a product of two quadratic factors in $R_3$ ---
and similar techniques can be used to discuss bifurcations of its solutions. Specifically, there can be a double root
if the discriminant of one of the factors vanishes (i.e. if $Q_2$ or $Q_3$ vanish), or if the factors have a common root.
The latter happens in the two cases 
\begin{equation}
  \left( (2\gamma_1\pm\gamma_4)R_1+2 \gamma_2\right)^2+ 4Q_3 \gamma_3\gamma_4  = 0\ .
\label{zzzz}\end{equation}  
As in Section 3, detecting beating transitions requires translating the initial conditions to the constants of
motion $R_1,Q_2,Q_3$ and checking up to $4$ conditions.

We have implemented this method for the CQ and SAT systems and found some satisfactory results which
we do not report here; in certain cases the results were better than those found using the method based on
canonical transformations. However there are numerous reasons to prefer the method based on canonical transformations. 
The two time method requires deciding how to explicitly introduce a small parameter and different ways
of doing this give different results. It also requires advance knowledge of the correct relative order of
magnitude of the oscillations around the fixed point and the deviation of the system parameters from their
resonance values. In general,
the algebraic manipulations required to implement the two time method, most of which
we have omitted in our account here, are substantially more complicated than those required for the method based
on canonical transformations; in particular the reduction of the system (\ref{Reqs}) 
to a single differential equation (\ref{R3eq}) is a surprise, that emerges from {\em ad hoc} manipulations,
whereas the parallel steps in the canonical formalism are standard, based on the integrability of the Hamiltonian
(\ref{H3}). Finally, from our numerical experiments it emerges that
while the results based on the vanishing of the discriminant
of one of the factors of the right hand side of (\ref{R3eq}) are good, the results based on conditions (\ref{zzzz})
are poor. 


\bibliographystyle{acm}
\bibliography{w}

\end{document}